\documentclass[12pt,reqno]{article}
\usepackage{arxiv}
\usepackage[letterpaper]{geometry}
\newif\ifclean
\cleantrue  
\usepackage{graphicx}
\usepackage{amsmath}
\usepackage{amssymb}
\usepackage{amsfonts}
\usepackage{subcaption}
\graphicspath{{{./}}}

\ifclean
\else
\usepackage{color}

\newcommand{\red}[1]{\textcolor{red}{{#1}}}
\newcommand{\caution}{\red{\bf Draft: \today. Do not distribute.}}
\fi

\usepackage{amsthm}
\newtheorem*{remark}{Remark}

\newcommand{\aref}[1]{App.\,\ref{#1}}
\newcommand{\fref}[1]{Fig.\,\ref{#1}}
\newcommand{\tref}[1]{Table\,\ref{#1}}
\newcommand{\eref}[1]{Eq.\,(\ref{#1})}

\newcommand{\sref}[1]{Sec.\!~\ref{#1}}

\newcommand{\cref}[1]{Ref.\,\cite{#1}}
\newcommand{\crefs}[1]{Refs.\,\cite{#1}}

\newcommand{\ie}{{\it i.e.}\!\, }
\newcommand{\eg}{{\it e.g.}\!\, }

\newcommand{\etal}{{\it et al.} }
\newcommand{\apriori}{{\it a priori} }

\newcommand{\order}{\operatorname{order}}

\newcommand{\tighteq}{\mkern1.5mu{=}\mkern1.5mu}
\renewcommand{\Re}{\operatorname{Re}}
\newcommand{\Abb}{\mathbb{A}}
\newcommand{\Ibb}{\mathbb{I}}
\newcommand{\Jbb}{\mathbb{J}}
\newcommand{\Cbb}{\mathbb{C}}
\newcommand{\Dbb}{\mathbb{D}}

\newcommand{\Obb}{\mathbb{O}}

\newcommand{\ab}{\mathbf{a}}

\newcommand{\eb}{\mathbf{e}}

\newcommand{\ub}{\mathbf{u}}
\newcommand{\vb}{\mathbf{v}}
\newcommand{\rb}{\mathbf{r}}

\newcommand{\nb}{\mathbf{n}}
\renewcommand{\sb}{\mathbf{s}}
\newcommand{\Ab}{\mathbf{A}}
\newcommand{\Bb}{\mathbf{B}}

\newcommand{\Eb}{\mathbf{E}}

\newcommand{\Gb}{\mathbf{G}}

\newcommand{\Lb}{\mathbf{L}}
\newcommand{\Mb}{\mathbf{M}}
\newcommand{\Ib}{\mathbf{I}}
\newcommand{\Rb}{\mathbf{R}}
\newcommand{\Sb}{\mathbf{S}}

\newcommand{\Xb}{\mathbf{X}}

\newcommand{\Bc}{\mathcal{B}}
\newcommand{\Gc}{\mathcal{G}}
\newcommand{\Ic}{\mathcal{I}}
\newcommand{\Oc}{\mathcal{O}}

\newcommand{\as}{\mathsf{a}}
\newcommand{\hs}{\mathsf{h}}
\newcommand{\fs}{\mathsf{f}}
\newcommand{\ms}{\mathsf{m}}
\newcommand{\rs}{\mathsf{r}}

\newcommand{\ys}{\mathsf{y}}
\newcommand{\zs}{\mathsf{z}}
\newcommand{\xs}{\mathsf{x}}
\newcommand{\ws}{\mathsf{w}}

\newcommand{\Ds}{\mathsf{D}}

\newcommand{\Ps}{\mathsf{P}}

\newcommand{\Vs}{\mathsf{V}}

\newcommand{\Ys}{\mathsf{Y}}

\newcommand{\Ws}{\mathsf{W}}
\newcommand{\bs}{\mathsf{b}}

\newcommand{\Rs}{\mathsf{R}}

\newcommand{\phib}{{\boldsymbol{\phi}}}
\newcommand{\varphib}{{\boldsymbol{\varphi}}}
\newcommand{\tr}{{\operatorname{tr}}}
\newcommand{\perm}{{\operatorname{perm}}}
\newcommand{\sym}{{\operatorname{sym}}}
\newcommand{\asym}{{\operatorname{asym}}}
\newcommand{\dev}{{\operatorname{dev}}}
\newcommand{\devsym}{{\operatorname{devsym}}}

\newcommand{\partialb}{{{\boldsymbol{\partial}}}}

\newcommand{\Conv}{{\operatorname{Conv}}}

\newcommand{\graph}{{\mathit{G}}}
\newcommand{\group}{{\mathcal{G}}}
\newcommand{\structuraltensor}{{\mathbb{A}}}
\newcommand{\modulustensor}{{\mathbb{C}}}
\newcommand{\basis}{{\mathcal{B}}}
\newcommand{\invariants}{{\mathcal{I}}}
\newcommand{\strain}{{\mathbf{E}}}
\newcommand{\stress}{\mathbf{S}}
\newcommand{\RE}{{R}}
\newcommand{\CG}[3]{\mathsf{C}^{(#1)}_{(#2,#3)}}

\title{\bf\Large Equivariant graph convolutional neural networks for the representation of homogenized anisotropic microstructural mechanical response}
\author{Ravi Patel, \\
Sandia National Laboratories,\\
Albuquerque, NM 87185
\And
Cosmin Safta, \\
Sandia National Laboratories,\\
Livermore, CA 94551
\And
Reese Jones\footnote{\tt rjones@sandia.gov} \\
Sandia National Laboratories,\\
Livermore, CA 94551
}

\begin{document}
\ifclean
\date{}
\else
\date{\caution}
\fi

\maketitle{}

\begin{abstract}
Composite materials with different microstructural material symmetries  are common in engineering applications where grain structure, alloying and particle/fiber packing are optimized via controlled manufacturing.
In fact these microstructural tunings can be done throughout a part to achieve functional gradation and optimization at a structural level.
To predict the performance of particular microstructural configuration and thereby overall performance, constitutive models of materials with microstructure are needed.

In this work we provide neural network architectures that provide effective homogenization models of materials with anisotropic components.
These models satisfy equivariance and material symmetry principles inherently through a combination of equivariant and tensor basis operations.
We demonstrate them on datasets of stochastic volume elements with different textures and phases where the material undergoes elastic and plastic deformation, and show that the these network architectures provide significant  performance  improvements.

\end{abstract}

\section{Introduction}

Symmetry preservation is a key to effective computational physics.
The success of symplectic integrators \cite{hairer2006structure,marsden2013introduction} in predicting long time dynamics is one notable example.
In fact, symmetries and other \apriori exact constraints reduce unneeded complexity in models and can promote stability.
In the field of constitutive modeling, the primary symmetry is expressed by the principle of \emph{frame invariance} \cite{truesdell2004non}, or \emph{objectivity}, which prescribes how a physical function must respond to changes in observer/coordinate frame.
Simply put, rotations of the tensorial inputs to a constitutive function must result in corresponding rotations of the output.
Now more commonly called \emph{equivariance}, at least in the scientific machine learning (SciML) literature, this principle allows for elegant simplification of the form of physical models, particularly for isotropic materials whose response to deformation is invariant to all (distance-preserving) rotations of their reference configurations.
For the subclass of materials that have pronounced anisotropy, a more limited invariance holds for the subset of rotations that are in their symmetry group.
For these materials all transformations of the reference configuration in the symmetry group of the material give the same response to any given loading.

There are a number of approaches to embed these symmetries in SciML models so that they hold by construction.
Tensor basis (TB) formulations \cite{ling2016machine} rely on classical tensor basis expansions (TBEs) from  representation theory going back to Finger \cite{finger1894potential}, Rivlin \cite{rivlin1955stress} and contemporaries \cite{smith1957anisotropic,spencer1962isotropic,boehler1987applications}.
This is still an active field in mechanics and continues with recent work \cite{olive2017minimal,desmorat2021minimal}.
Based on Rivlin's work, Ling \etal \cite{ling2016machine} invented the \emph{Tensor Basis Neural Network} (TBNN) and utilized it to model crystal response, and fluids applications.
The formulation is adaptable to multiple inputs \cite{ling2016machine,jones2018machine} and  the often cited monograph of Zheng \cite{zheng1994theory} provides the joint invariants and tensor basis generators for a variety of modeling scenarios.
Jones \etal extended the TBNN paradigm to formulations for phenomenological plasticity  \cite{jones2018machine} and to inelastic materials with microstructure \cite{jones2022neural}.
By the same method anisotropy can be embedded through the addition of structure tensors characterizing the material symmetry group, which can be learned from data as in the work of Fugh \etal \cite{fuhg2022learning} .
The basic formulation is suitable for other ML representations, for example Gaussian Processes \cite{frankel2020tensor,fuhg2022physics}.
A recent contribution \cite{fuhg2023stress} surveys the variety of TBNN formulations for hyperelasticity.

In contrast to the TBNN approach, equivariant neural networks focus on the combination of information that is not at the same material location, as in convolution of field data.
This aspect of the equivariance principle is particularly germane to materials with microstructure, where the principle involves the rotations of finite size representative samples not just tensorial data at material points.
Pixel-based convolutional neural networks can learn this symmetry inexactly, given enough data.
In fact augmentation has been used to exploit this route, for example
Dielemann \etal \cite{dieleman2016exploiting} augmented the convolutional layers of a CNN with the orthogonal transformation accessible on a grid, \ie  $\pi/4$ rotations and inversions.
Graph convolutional neural networks (GCNN) \cite{kipf2016semi,frankel2022mesh} have a degree of embedded equivariance due to the permutational invariance of their convolution kernels, \ie the kernel is not a function of the position.
Along these lines, there have been approaches based on constrained kernel construction.
Worrall \etal \cite{worrall2017harmonic} developed filters based on circular harmonics and Chidester \etal \cite{chidester2018rotation} used a discrete Fourier transform to embed rotational invariance in convolutional filters.

General strict equivariance for convolutional NNs began with theoretical developments.
Cohen and Welling \cite{cohen2016group} outlined the mathematics whereby convolutional networks will be equivariant with respect to any group.
Then Kondor and Trivedi \cite{kondor2018generalization} proved that convolutional structure is  necessary and sufficient for equivariance to the action of a compact symmetry group using concepts from abstract algebra.
Finzi \etal \cite{finzi2020generalizing} provided a significant extension of the Cohen and Welling treatment of small, discrete symmetry groups to continuous (Lie) groups, such as the special orthogonal group SO(3).
Practical implementations of equivariant neural networks (EqvNNs) followed.

A particular formulation of equivariant convolutions using spherical harmonic expansion and the mathematics of tensor product spaces orchestrated by the Clebsch-Gordon algebra was developed by Thomas \etal \cite{thomas2018tensor} and followed by Batzner \etal \cite{batzner20223} and Schmidt \etal \cite{smidt2021euclidean}.
This \emph{Tensor Field Network} (TFN) was created with molecular interactions in mind and was applied to point cloud data.
This development was quickly followed up with applications to continuous fields.
For instance, Sun and coworkers have been particularly active in this field.
Vlassis \etal \cite{vlassis2020geometric} used it in the context of hyperelasticity, Heider \etal \cite{heider2020so} employed it in modeling phenomenological plasticity, and Cai \etal \cite{cai2023equivariant} applied it to represent permeability fields.

In this work we adapt the equivariant TFN to the task of representing the response of polycrystalline materials.
Our contributions are:
(a) hybridizing the EqvNN with a TBNN for the homogenization task,
(b) demonstrating the reuse weights on edges is accurate and effective in reducing overparameterization,
(c) showing how pruning can combat the combinatorial explosion of the spherical harmonic expansion and provide generalization,
and
(d) extension of the TB-EqvNN to evolving processes through novel recurrent and ordinary differential equation neural network implementations.
The proposed framework is illustrated by multiple architectures that are show to be effective at the task of homogenization.

In the next section we introduce homogenization in the setting of anisotropic constituents, starting with  representation theory for homogeneous materials.
We also pose the challenge of finding a compact representation for an aggregate of multiple constituents with different symmetries.
In \sref{sec:data}, we review the data generating model for the polycrystalline exemplar and provide details of the diverse datasets we use for the following demonstrations.
\sref{sec:arch} provides the details of the proposed tensor basis, equivariant graph convolutional neural network architecture that can be employed as representation for the homogenized response of polycrystals and similar materials.
This section also includes the mathematics of irreducible representations (IRs), spherical harmonic expansion and how these interplay with classical tensor basis representations.
\aref{app:irrep} provides additional exposition of the fundamentals of IRs.
Then \sref{sec:results} provides demonstrations of the efficacy of the TB-EqvGCNN for crystal elasticity and crystal plasticity exemplars.
Finally, we summarize the findings and give directions for future work in \sref{sec:conclusion}.

\section{Homogenization of anisotropic aggregates} \label{sec:homogenization}

In this section, first we introduce the fundamentals of anisotropy for solids, then the homogenization problem and the representation challenge that follows.
Throughout the following, we use $a$ to index basis elements in tensor basis expansions, $I$ to index cells in discretized fields, and $i$ to index spatial tensor components.

\subsection{Symmetry and structural tensors}

A \emph{structural tensor} $\structuraltensor$ is a tensor that characterizes a point group symmetry, i.e., a finite subgroup of the proper rotations, $\group \subset \textrm{SO(3)}$, via the property of invariance to group action
\begin{equation} \label{eq:structural_tensor}
\Gb \boxtimes \structuraltensor = \structuraltensor \ \ \forall \ \Gb \in \group_\structuraltensor ,
\end{equation}
so that an anisotropic response function $\hat{\stress}(\strain)$,
\begin{equation} \label{eq:anisotropic_function}
\hat{\Sb}(\Gb \boxtimes \Eb) = \hat{\Sb}(\Eb)
\end{equation}
can be represented with an isotropic function
\begin{equation}
\stress = \hat{\stress}(\strain, \structuraltensor)
\end{equation}
of two arguments, the original input $\strain$ and the structural tensor $\structuraltensor$.
Here $\Gc_\structuraltensor$ is the symmetry group and $\boxtimes$ is the Kronecker product  \ie
\begin{equation}\label{eq:kronecker_product}
\Gb \boxtimes \structuraltensor
= \Gb \boxtimes \sum_{i,j,..} [\structuraltensor]_{ij...} \eb_i \otimes \eb_j \otimes ...
= \sum_{i,j,..} [\structuraltensor]_{ij...} \Gb \eb_i \otimes \Gb \eb_j \otimes ...
\end{equation}
follows from the Rychlewski-Zhang theorem \cite{zhang1990structural},
which is sometimes called the {\it isotropization} theorem.
The operation $\Gb \boxtimes$ defines group action of $\Gb$ for (Cartesian) tensors of any order.

In this work, we concentrate on the stress response $\stress$ due to elastic strain $\strain$.
In this case, equivariance is defined as
\begin{equation} \label{eq:equivariance}
\Gb \boxtimes \stress(\strain,\structuraltensor) = \stress( \Gb \boxtimes \strain,  \Gb \boxtimes \structuraltensor)  \quad \Gb \in \text{SO(3)}
\end{equation}
which requires that a change of basis of the inputs results in the corresponding rotation of the output.
This is also the definition of an isotropic function.
A key result of \eref{eq:structural_tensor} is
\begin{equation} \label{eq:material_symmetry}
\stress(\strain, \Gb \boxtimes \structuraltensor) =  \stress(\strain, \structuraltensor) \ \forall \ \Gb \in \group_\structuraltensor
\end{equation}
\ie
changing the reference orientation with a symmetry operation $\Gb \in \group_\structuraltensor$ leads to the same response.
In other words  rotating the material by $\Gb$ but keeping the imposed strain $\strain$ the same results in same stress $\stress$.
Given the structural tensor  $\structuraltensor$, the stress $\stress$ can then be represented with a tensor basis \cite[Ch.3]{boehler1987representations}:
\begin{equation} \label{eq:tbnn}
\stress = \stress(\strain, \structuraltensor) = \sum_a \hat{c}_a(\invariants) \Bb_a ,
\end{equation}
formed from the product of coefficients $c_a$, which are functions of the joint scalar invariants $\invariants$, and tensor basis $\basis=\{\Bb_a\}$.
This assumes the appropriate invariants and basis for the inputs $\{\strain,\structuraltensor\}$ can be found  \cite{boehler1987representations,zheng1994theory}.

\begin{remark}
We have been non-specific about the particular stress measure.
Both the Cauchy stress and the second Piola-Kirchhoff stress are symmetric and have the similar representations; however, the second Piola-Kirchhoff has the advantages of direct connection with a potential and being in the same frame as the structure tensor.
While the second Piola-Kirchhoff stress is invariant with respect to \emph{superposed rigid motions} \cite{green1979note,murdoch2003objectivity} in the current configuration, it still needs to respect a change of basis in the reference, known as \emph{material frame indifference} \cite{truesdell2004non}.
Hence, we associate $\stress$ with second Piola-Kirchhoff stress for the remainder of this work.
\end{remark}

\subsection{Homogenization}
We focus on the prediction of the evolution of the average of stress $\bar{\stress}$ with strain $\strain(\Xb,t)$ in a sample region $\Omega$:
\begin{equation} \label{eq:phi_problem}
\bar{\stress}(t) \equiv \frac{1}{V} \int_\Omega \stress\left( \strain(\Xb,t), \phib(\Xb) \right) \, \mathrm{d}^3X \ ,
\end{equation}
with volume $V$ composed of anisotropic component materials characterized by a microstructural field $\phib(\Xb)$ observed at time $t=0$.
Here $\Xb$ is the reference position of the material and $t$ is time.
Formally we exchange this formulation, based on, for example, texture orientation angles $\phib(\Xb)$ \cite{frankel2019oligocrystals,frankel2019evolution,frankel2020prediction}, for a formulation based on the structure tensors $\structuraltensor(\Xb)$ of each point $\Xb$:
\begin{equation} \label{eq:H_problem}
\bar{\stress}(t) \equiv \frac{1}{V} \int_\Omega \stress\left( \strain(\Xb,t), \structuraltensor(\Xb) \right) \, \mathrm{d}^3X
\end{equation}
This shift is important since it allows access of representation theory \cite{spencer1958theory,boehler1987representations,zheng1994theory,rivlin1997stress} and clarifies the equivariance requirements, as in \eref{eq:equivariance}.
We assert the isotropization theorem holds for this case so that the function $\bar{\stress}$ has to be equivariant to SO(3):
\begin{equation}
\Gb \boxtimes \bar{\stress} \equiv \frac{1}{V} \int_\Omega \stress\left(\Gb \boxtimes \strain, \Gb \boxtimes \structuraltensor \right) \, \mathrm{d}^3X \quad \forall \Gb \in \text{SO(3)} \ .
\end{equation}
Furthermore the analog of \eref{eq:material_symmetry} holds by virtue of \eref{eq:structural_tensor}.

To obtain the stress field  $\stress(\strain(\Xb,t), \structuraltensor(\Xb))$ the compatible strain field $\strain(\Xb,t)$ needs to be solved for given the imposed strain, which is prescribed by the boundary or the average strain $\bar{\strain}(t)$.
The goal of the NN models described in \sref{sec:arch} is to predict the mean stress $\bar{\stress}$ given  only: (a) the initial microstructure $\structuraltensor(\Xb)$, and (b) the external, applied strain $\bar{\strain}(t)$, in effect bypassing the boundary value solve.

\subsection{Representation problem} \label{sec:representation_problem}
The representation problem is to find the model, $\bar{\stress}$, as a function of the external loading $\bar{\strain}(t)$ and the microstructural anisotropy field $\structuraltensor(\Xb)$ to the sample average stress $\bar{\stress}(t)$.
The fact that many microstructures, such as grains of a polycrystal or phases of an alloy, are composed of a collection of distinct subdomains, allows us to decompose $\Omega$ into finite number of subdomains with constant properties.
Furthermore we need to discretize the domain $\Omega$ into mesh of cells $\cup_K \Omega_K = \Omega$ each with a constant $\{ \structuraltensor_K \}$ and an (invariant) distance graph $\graph$ that encodes the spatial relationships of the cells.
With a finite collection of homogeneous subdomains, the problem involves approximating
\begin{equation} \label{eq:A_problem}
\bar{\stress}(t) \equiv \frac{1}{V} \int \stress\left( \strain(\Xb,t), \structuraltensor(\Xb) \right) \, \mathrm{d}^3X
\approx
\frac{1}{V} \sum_K \stress\left( \strain_K(t), \structuraltensor_K, \graph \right) \Delta\!V
\approx
\tilde{\stress}(\bar{\strain}(t),\{ \structuraltensor_K \}, \graph)
\end{equation}
while preserving
\begin{equation} \label{eq:A_equivariance}
\Gb \boxtimes \tilde{\stress}(\bar{\strain}(t),\{ \structuraltensor_K \}, \graph ) =  \tilde{\stress}\left( \Gb \boxtimes \bar{\strain}(t), \{ \Gb \boxtimes \structuraltensor_K \}, \graph \right) \quad \forall \Gb \in \text{SO(3)}
\end{equation}

To our knowledge there is no exact theory that covers this case, \ie  where we are trying to combine the effects of a field or multiple subcomponents which do not have the same symmetries (which we call \emph{heteroanisotropy}) or are rotations of the same symmetry (which we call \emph{homoanisotropy}).
Certainly classical approximate mixture rules \cite{nemat2013micromechanics,mura2013micromechanics} are applicable, such as the Hill-Voigt-Reuss \cite{reuss1929berechnung,hill1952elastic,voigt2014lehrbuch} and the Taylor-Sachs \cite{sachs1928plasticity,taylor1938plastic} estimators, but these can be inaccurate \cite{frankel2019oligocrystals}.

\begin{remark}
The Hill average of constant strain (Voigt) and constant stress (Reuss) assumptions for the modulus $\Cbb$:
\begin{equation} \label{eq:hill_modulus}
\bar{\Cbb}_\text{Hill} = \frac{1}{2} \left(
\sum_K \nu_K \Rb_K \boxtimes \Cbb
+ \left[ \sum_K \nu_K \Rb_K \boxtimes \Cbb^{-1} \right]^{-1}
\right)
\end{equation}
can be an adequate estimator at small strains \cite{frankel2019oligocrystals} where the grain contributions are more or less independent.
Here $\nu_K = V_K/V$ are the volume fractions of the grains.
This mean mixture modulus tensor can be used to estimate stress with a St. Venant model: $\bar{\Sb} = \bar{\Cbb} \Eb$ and also give an expectation of the symmetries of polycrystalline aggregates.
Note that mixture estimators of this type ignore continuity/compatibility and relaxation of the displacement field.

\end{remark}

\begin{remark}
A few approaches to the representation problem beyond simple mixture rules are obvious but flawed.

One approach is to form a TB representation of using all the joint invariants and basis generators for $\{ \structuraltensor_K \}$, for example for 4th order $\structuraltensor_K$:
\begin{eqnarray}
\stress = \bar{\stress}(\bar{\strain},\{ \structuraltensor_I \}) &=& \hat c_0 \Ib + \hat c_1 \bar \strain + \hat c_2 \bar \strain^2 +  \\
&& \sum_I \left[
\hat c^{(1,1)}_{I} \sym \structuraltensor_I \bar \strain  +  \ldots \right] + \nonumber \\
&& \sum_{I,J} \left[
\hat c^{(1,2)}_{IJ} \sym \structuraltensor_I \bar \strain \structuraltensor_J +
\hat c^{(2,2)}_{IJ} \sym \structuraltensor_I \bar \strain^2 \structuraltensor_J +  \ldots \right] \nonumber
...
\end{eqnarray}
where the coefficients $\hat{c}_{I\ldots}$ are (NN) functions of all the scalar invariants.
Clearly this is cumbersome since the permutations in the joint invariants grows rapidly with the number of subdomains and the expansion ignores the spatial proximity encoded in $\graph$.

Another approach is to form TB representation using a single representative structure tensor $\bar{\structuraltensor}$, for example for a 4th order $\structuraltensor$:.
\begin{eqnarray}
\stress &=& \bar{\stress}(\bar{\strain},\bar{\structuraltensor}) = \sum_a \hat{c}_a(\invariants) \Bb_a \\
&=& \hat c_0 \Ib + \hat c_1 \bar \strain + \hat c_2 \bar \strain^2 +
\hat c_3 \,  \sym \bar \strain   \bar{\structuraltensor} +
\hat c_4 \,  \sym \bar \strain^2 \bar{\structuraltensor} + ... +
\hat c_8 \,  \sym \bar \strain^2 \bar{\structuraltensor}^2 \nonumber
\end{eqnarray}
This has the advantage that a single material tensor basis expansion can be employed; however, a secondary model $\bar{\structuraltensor} = \bar{\structuraltensor}(\{\structuraltensor_K\})$ needs to be constructed to determine $\bar{\structuraltensor}$ and \apriori assumptions, such as  $\bar{\structuraltensor}$ is the same rank as $\structuraltensor_K$, are likely restrictive \cite{moakher2006closest}.
The fact that even an aggregate of homoanisotropic subdomains does not result in a symmetry from the same or with a structural tensor of the same order is illustrated by the isotropic limit, where, for example,  a large collection of cubic crystals, with 4th order structural tensors, tends to isotropy, which has the 2nd order identity as its characteristic tensor.

These issues motivate a mixture of tensor basis representations and equivariant convolution described in \sref{sec:arch}.
\end{remark}

\section{Polycrystal data} \label{sec:data}

Polycrystalline elastoplasticity is a canonical exemplar of the behavior of aggregates of constituents with different material symmetries.
Polycrystalline aggregates are ubiquitous in the materials science of metals, ceramics, rocks and ice, and at the center of many efforts to upscale micromechanics to large scale plasticity phenomenology \cite{taylor1934mechanism,kroner1961plastic,bishop1951xlvi,bishop1951cxxviii,mandel1965generalisation,dawson2000computational,roters2010overview}.

Modeling the stress response of polycrystalline materials is important in simulating these materials at a larger scale.
In polycrystalline materials, microstructure is a composition of subdomains each with an anisotropic response.

The datasets are comprised of polycrystalline stochastic volume elements (SVEs)  where boundary conditions are used to evoke nominally homogeneous deformations.
The SVEs are finite size samples which display response variability sample to sample.
The quantity of interest in our study is the sample average stress.

\subsection{Elasticity} \label{sec:crystal_elasticity}
A polycrystal composed of face centered cubic (FCC) crystals is a particularly common example in structural metals.
Each FCC crystal in the aggregate has the point symmetry group $\Oc_h$ and a 4th order structure tensor \cite{zheng1994theory}:
\begin{equation} \label{eq:cubic_structural_tensor}
\structuraltensor = \Obb_h \equiv
\sum_{i=1}^3 \eb_i \otimes \eb_i \otimes \eb_i \otimes \eb_i
= \delta_{ij} \delta_{jk} \delta_{kl} \delta_{li}  \eb_i \otimes \eb_j \otimes \eb_k \otimes \eb_l \ ,
\end{equation}
if the symmetry axes are aligned with the Cartesian axes $\{ \eb_i \}$.
Any other orientation can be obtained via rotation $\Rb = \Rb(\phib) \in \text{SO(3)}$
\begin{equation}
\Rb \boxtimes \structuraltensor
= \sum_{i=1}^3 \Rb \eb_i \otimes \Rb \eb_i \otimes \Rb \eb_i \otimes \Rb \eb_i
= \sum_{i=1}^3 \rb_i \otimes \rb_i \otimes \rb_i \otimes \rb_i
\end{equation}
to alignment with local texture/orientation $\phib(\Xb)$.
The associated elastic modulus tensor
\begin{equation}
\modulustensor \equiv \partialb_\strain \stress  = \lambda \Ibb + 2 \mu \Jbb + \alpha \structuraltensor
\end{equation}
is composed of two isotropic 4th order tensors \cite{HASHIGUCHI20201},
\begin{eqnarray}
\Ibb &=& \Ib \otimes \Ib =  \delta_{ij} \delta_{kl} \eb_i \otimes \eb_j \otimes \eb_k \otimes \eb_l \\
\Jbb &=& \frac{1}{2} ( \delta_{ik} \delta_{jl} +  \delta_{il} \delta_{jk} ) \eb_i \otimes \eb_j \otimes \eb_k \otimes \eb_l ,
\end{eqnarray}
and the cubic structural tensor $\structuraltensor$ \cite{tu1968decomposition}.
The isotropic tensors  $\Ibb$ and $\Jbb$ are invariant to SO(3), which is evident since pairs of $\Rb$ are summed and $\Rb^T \Rb = \Rb \Rb^T = \Ib$.
In terms of the more familiar cubic modulus constants: $C_{11} \equiv [ \Cbb ]_{iiii}  = \lambda+2\mu+\alpha$, $C_{12} \equiv [ \Cbb ]_{iijj}  = \lambda$, and $C_{44} \equiv [ \Cbb ]_{ijij}  = 2\mu$.

For a single crystal, Kambouchev \etal \cite{kambouchev2007polyconvex} provides a general polyconvex hyperelastic model with cubic symmetry.
The elastic potential $\Phi$,
\begin{equation} \label{eq:kam_invar}
\Phi = \Phi(\strain,\structuraltensor)
= \Phi(\underbrace{\tr \strain, \tr \strain^*, \det \strain}_{\invariants_\text{iso}}, \underbrace{\structuraltensor:\strain\otimes\strain, \structuraltensor:\strain \otimes \strain^2, \structuraltensor: \strain^2 \otimes \strain^2}_{\invariants_\structuraltensor}) \ ,
\end{equation}
takes the scalar invariants $\invariants = \invariants_\text{iso} \cup \invariants_\structuraltensor$ as inputs, where $ \invariants_\structuraltensor$ are the joint invariants of $\bar{\Eb}$ and $\Abb$.
Here $\Eb^*=\det(\Eb)\Eb^{-1}$ is the adjugate of $\Eb$ and three traces of products of the structural tensor $\structuraltensor$ and the strain $\strain$ are the anisotropic scalar invariants, $\invariants_\structuraltensor$.
Note, an equivalent set of invariants are $\invariants_\text{iso} = \tr \strain, \tr \strain^2, \tr \strain^3$.
For the cubic structure tensor  $\Abb = \Obb_h$ the anisotropic invariants can be reduced to
\begin{equation}
\invariants_\structuraltensor = \left\{
\sum_i (\rb_i \cdot \Eb \rb_i)^2,
\sum_i (\rb_i \cdot \Eb \rb_i) (\rb_i \cdot \Eb^2 \rb_i),
\sum_i (\rb_i \cdot \Eb^2 \rb_i)^2
\right\}
\end{equation}
A corresponding tensor basis can be obtained from the derivatives of the invariants:
\begin{equation} \label{eq:kam_basis}
\basis = \{ \partial_\strain \invariants_i \} = \{ \Bb_1\tighteq \Ib, \Bb_2\tighteq \strain, \Bb_3\tighteq \strain^* \} \cup \{ \Bb_4, \Bb_5, \Bb_6 \}
\end{equation}
where
\begin{eqnarray}
\Bb_4 &=&
2 \structuraltensor_{IJPQ}  \strain_{PQ} \eb_I \otimes \eb_J \\
&=&  \sum_I 2 \left( \rb_I \cdot \strain \rb_I \right) \rb_I \otimes \rb_I \nonumber \\
\Bb_5 &=&
\bigl[
\structuraltensor_{IJPQ} \strain^2_{PQ}  +
\strain_{MN} ( \structuraltensor_{MNIQ} \strain_{JQ} + \structuraltensor_{MNJQ} \strain_{IQ} ) \bigr] \eb_I \otimes \eb_J \\
&=& \sum_I ( \rb_I \cdot \strain^2 \rb_I + 2 (\rb_I \cdot \strain \rb_I)^2 ) \rb_I \otimes \rb_I \nonumber \\
\Bb_6 &=&
2 \strain^2_{MN} ( \structuraltensor_{MNIQ} \strain_{JQ} + \structuraltensor_{MNJQ} \strain_{IQ} ) \eb_I \otimes \eb_J  \\
&=& 4 \sum_I ( (\rb_I \cdot \strain^2 \rb_I) (\rb_I \cdot \strain \rb_I) ) \rb_I \otimes \rb_I \ . \nonumber
\end{eqnarray}
for $\structuraltensor$ satisfying the same symmetries as the elastic modulus tensor: $\structuraltensor_{IJPQ} = \structuraltensor_{JIPQ} = \structuraltensor_{PQIJ} = \structuraltensor_{PQJI}$.
This basis spans the full output space of symmetric tensors, unless the eigenvectors of $\Eb$ are aligned with the cubic axes $\{\rb_I\}$ in which case the basis is still sufficient to represent $\stress$.
Note the scalar invariants has all the traces of the tensor basis, with $\det \Eb$ substituted for the constant $\tr \Ib = 3$.
Also note $\strain^2_{PQ}$ is the $PQ$-th component of $\strain^2$ (not the square of the component).
The appendix of \cref{kambouchev2007polyconvex} provides derivatives of invariants with respect to the strain measure.
See also \cref{olive2017minimal} and \cref{desmorat2021minimal} for decompositions of the elasticity tensor and bases for elastic symmetry classes.

The stress of a polycrystalline SVE sample is obtained by solving for a continuous displacement field that is consistent with prescribed kinematic boundary conditions.
We use the finite element method to generate these solutions.

\subsection{Plasticity}
The previous section gives an overview crystal elasticity in a finite strain; however, metal crystals typically exhibit plasticity at moderate strains.
To generate plastic response data we used a common elastoviscoplastic model \cite{bishop1951xlvi,bishop1951cxxviii,frankel2019predicting}.

In each crystal, plastic flow
\begin{equation}
\Lb_p = \sum_\alpha \dot{\gamma}_{\alpha} \sb_\alpha \otimes \nb_\alpha
\end{equation}
could occur on any of the slip planes defined by the slip plane normal $\nb_\alpha$ and slip direction $\sb_\alpha$.
Here $\Lb_p$ is the plastic velocity gradient and $\dot{\gamma}_{\alpha}$ is the slip rate.
A power-law driven by the shear stress $\tau_\alpha$ resolved on slip system $\alpha$ governs the slip rate:
\begin{equation}
\dot{\gamma}_{\alpha}=\Gamma\left|\frac{\tau_{\alpha}}{g_{\alpha}}\right|^{m-1}\tau_{\alpha} \ .
\end{equation}
and in this model the slip resistance $g_\alpha$ evolves according to \cite{Kocks1976, mecking1976hardening}
\begin{equation}
\dot{g}_\alpha = (H-R\, g_\alpha) \sum_\alpha |\dot{\gamma}_\alpha| \ ,
\end{equation}
where $H$ is the hardening modulus and $R$ is the recovery constant.
See \cref{jones2018machine} for more details.

\subsection{Datasets} \label{sec:datasets}

We created two classes of datasets to demonstrate the NN architectures described in the next section:
(a) a comprehensive sampling of polycrystalline strain states for a variety of polycrystal types and realizations,  and
(b) a sampling of polycrystalline realizations undergoing plasticity in unaxial loading.
Realizations had between 4 and 26 grains which were discretized over 25$^3$ element meshes.

\begin{figure}[htb!]
\centering
\includegraphics[width=0.6\textwidth]{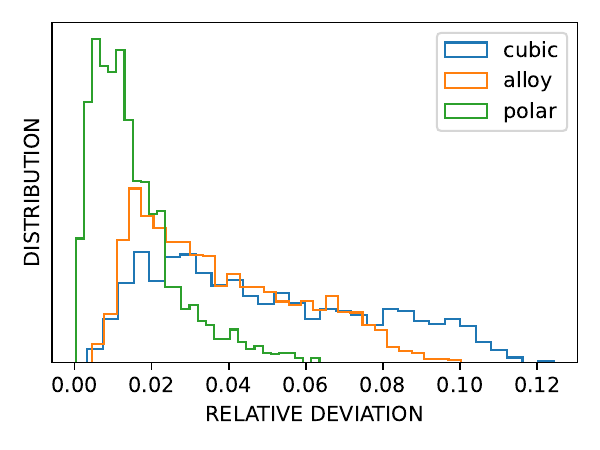}
\caption{Polycrystal: relative deviation of the stress from the Hill estimate {$\| {\bar{\Sb}-\bar{\Sb}_\text{Hill}} \| / \| \bar{\Sb} \|$} for the three elastic datasets.}
\label{fig:mixture_deviation}
\end{figure}

The elastic datasets included: (1) a \emph{homoanisotropic} cubic case with the properties of Fe (FCC steel) and uniform texture, (1) a  \emph{heteroanisotropic} binary alloy of cubic (Cu) and tetragonal (Sb) in a 9:1 ratio with uniform texture, and (3) a cubic Fe case with polarized orientation/texture.
All relevant material properties are given in \tref{tab:material_properties}.
For each of these datasets 2200 realizations were created, 1000 for training, 200 for in-training testing, and 1000 for independent, post-training validation.
Convex polyhedral grains in a unit cube were were assigned using a Voronoi tesselation.
A Karhunen-Lo\`{e}ve expansion (KLE) with a correlation length of 0.2 unit cells  was used to impart spatial correlation in the rotations $\Rb_K$ of neighboring crystals in the aggregate.
For the heteroanisotropic polycrystalline dataset we assigned some grains to be cubic with structural tensor $\Obb_h$ given in \eref{eq:cubic_structural_tensor} and others with the  tetragonal structural tensor:
\begin{equation} \label{eq:tetragonal_structural_tensor}
\Dbb_4 = \sum_{i,j,k,l=1}^2 \Re(\imath^{(i+j+k+l)}) \eb_i \otimes \eb_j \otimes \eb_k \otimes \eb_l  \ ,
\end{equation}
which differs from the cubic $\Obb_h$ structure tensor since the structure and response along one orthogonal axis is different than others in a tetragonal crystal.
Also note this structure tensor is not completely symmetric, unlike $\Obb_h$.
For the uniform texture cubic and mixed cubic/tetragonal data, the KLE generated a field of 4 Gaussian random variables which were transformed to samples on the unit 3-sphere by normalizing the 4-vector at every sample.
For the polar cubic dataset the KLE generated 2 Gaussian random variables with standard deviation 0.1 $\pi$ radians which were used to generate pole vectors deviating from a given direction and the rotations about these poles where sampled from a uniform distribution spanning 0 to 2 pi radians.
The field of structure tensors was obtained via $\Abb_K = \Rb_K \boxtimes \Abb^0_K$ where $\Abb^0_K$ is a structure tensor sampled from the selected distribution  in its canonical orientation.
For all elastic cases, the applied strains were generated from uniform samples the components of the displacement gradient in the hypercube $[-0.01,0.01]^9$ and the polycrystals were equilibrated to these nominal applied strains.
\fref{fig:mixture_deviation} shows that most of the dataset deviates significantly from the Hill estimate $\bar{\Sb}_\text{Hill} = \bar{\Cbb}_\text{Hill} \bar{\Eb}$ with $ \bar{\Cbb}_\text{Hill}$ given by \eref{eq:hill_modulus}.

\begin{figure}[htb!]
\centering
\includegraphics[width=0.24\textwidth]{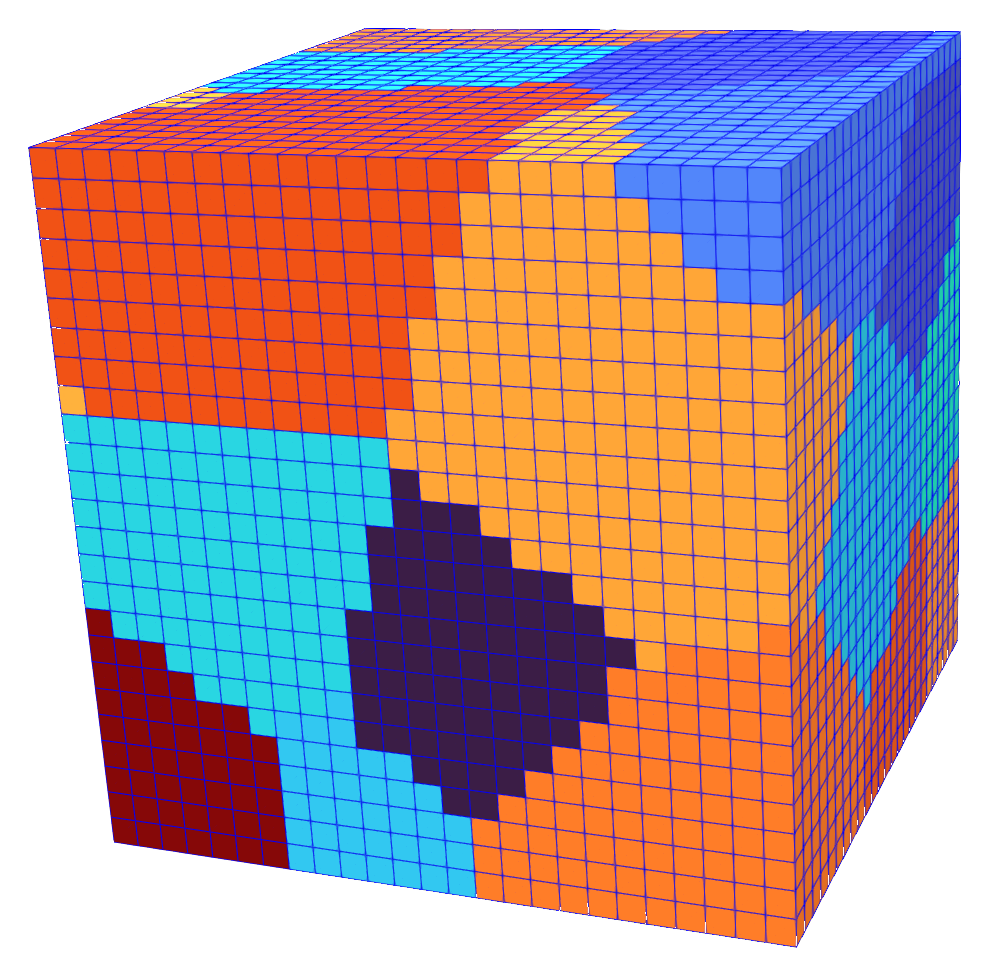}
\includegraphics[width=0.24\textwidth]{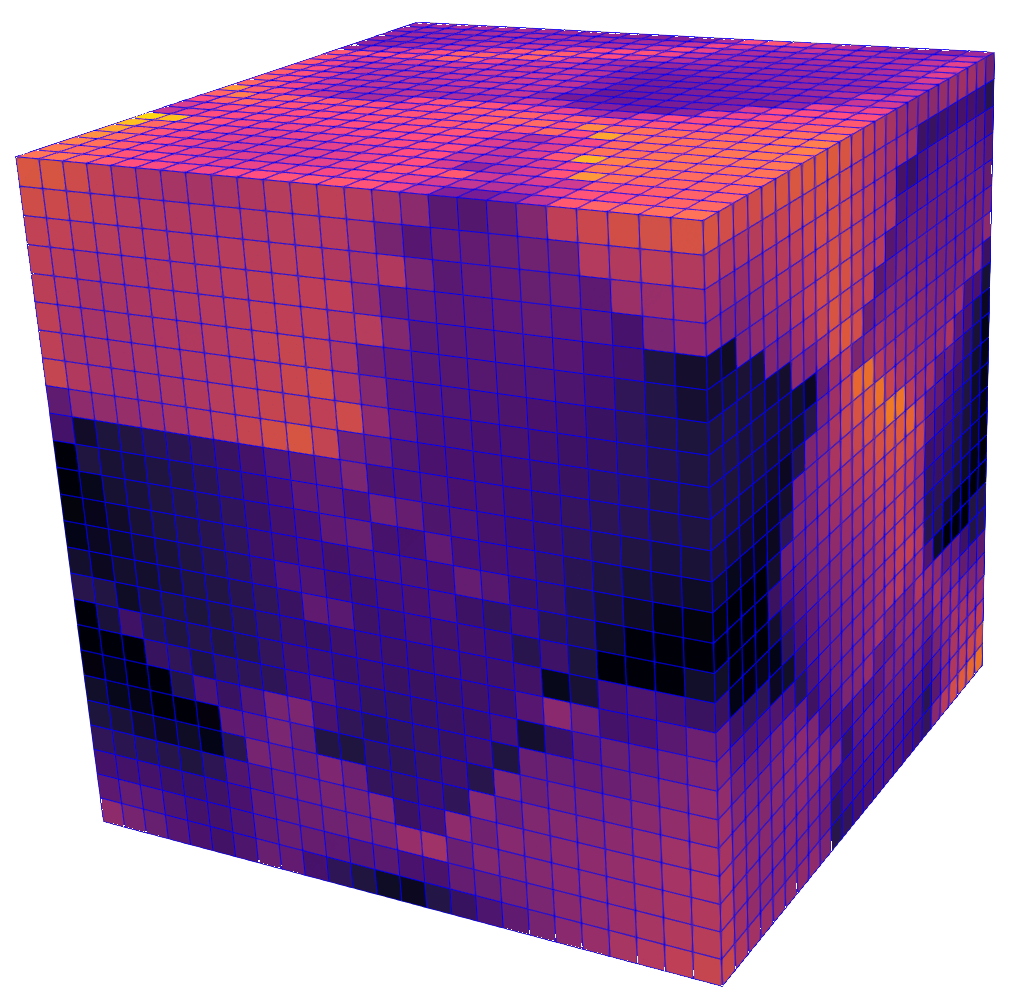}
\includegraphics[width=0.24\textwidth]{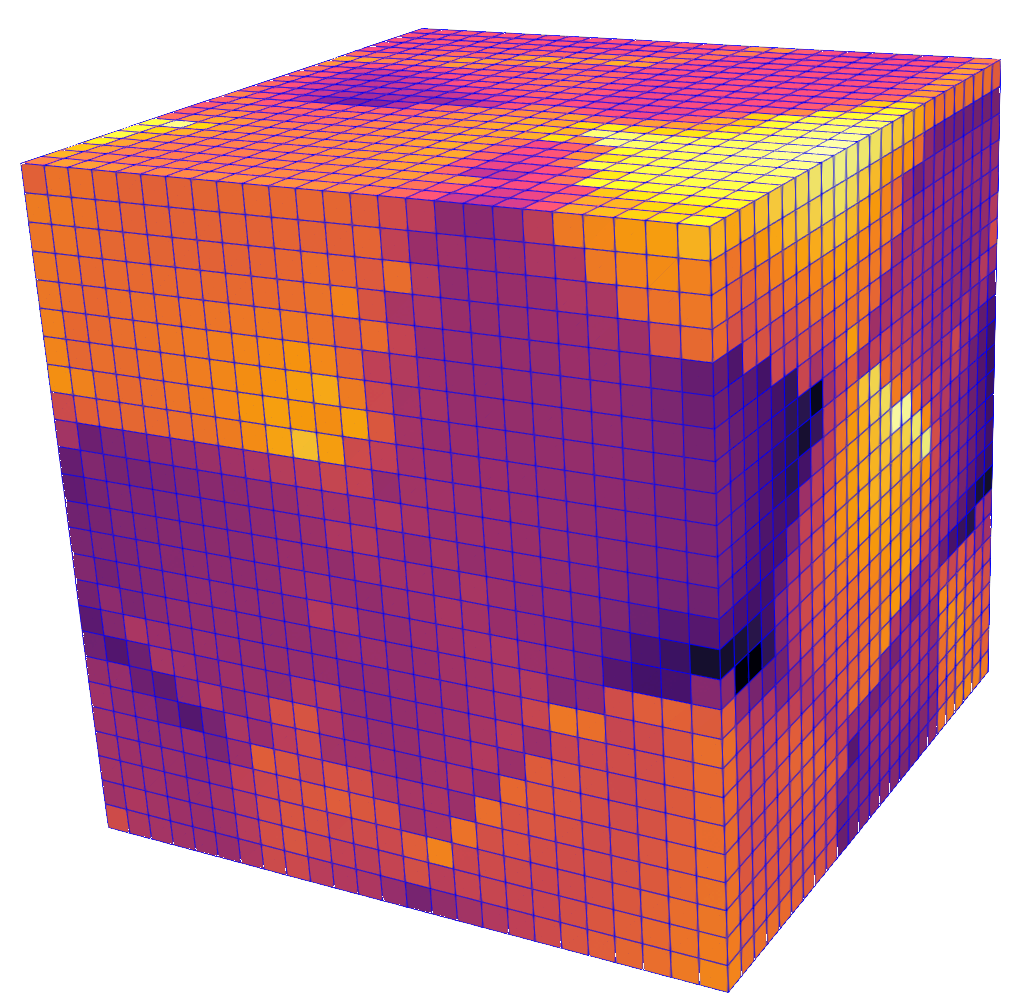}
\includegraphics[width=0.24\textwidth]{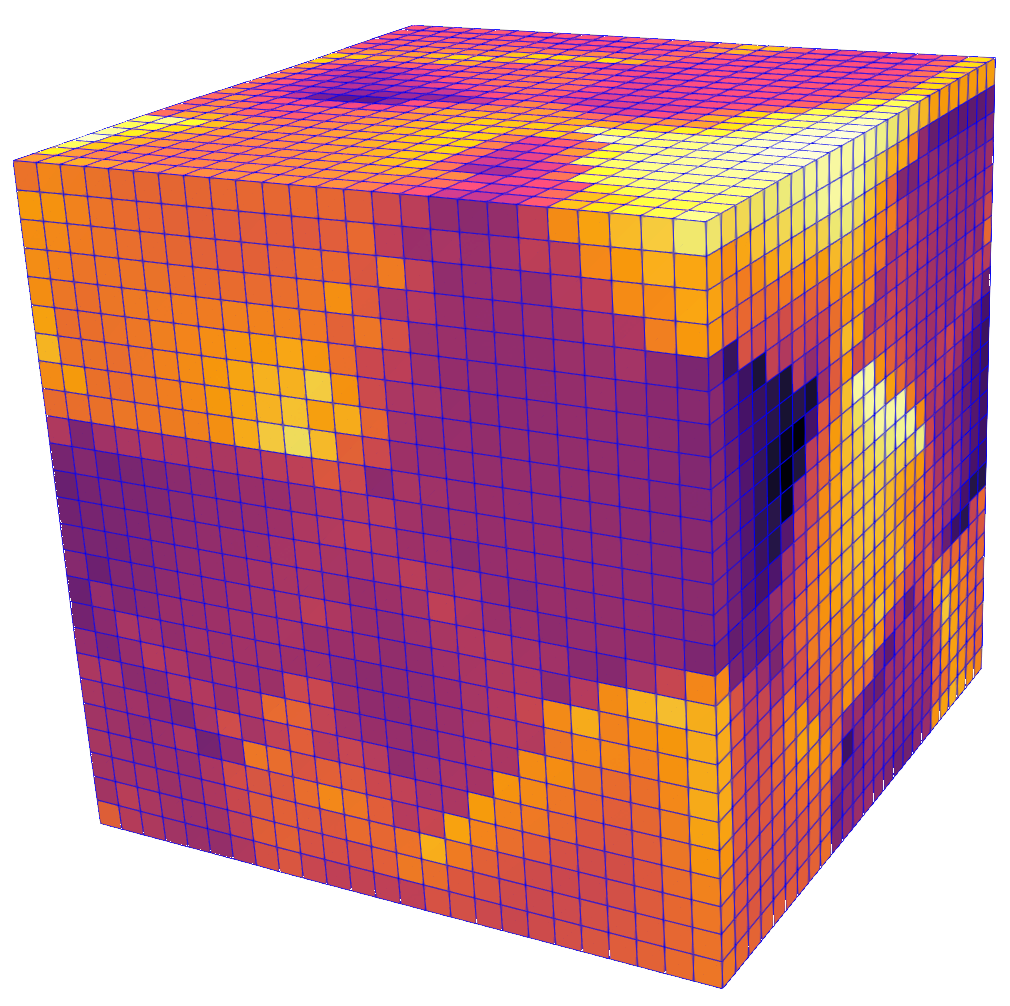}
\caption{Polycrystal (left to right panels): first Euler angle of orientation vector field $\phib$, stress $\stress$ states: elastic (0.1\%), transition (0.2\%), plastic (0.4\%) (colored by $\sigma_{11}$ with the same scale for all three panels [dark: $<$ 100 MPa, light: $>$ 400 MPa]).
}
\label{fig:stress_field}
\end{figure}

For the crystal plasticity dataset we used parameters representative of steel, refer to \tref{tab:material_properties}, and solved the equilibrium equations for tensile ramp-loading boundary conditions.
Using the same generation process, 7,000 samples were created with a different arrangement of FCC grains; 5,400 were used for training, 600 for in-training testing, and a 1000 for post-training validation.
Boundary conditions sufficient to effect the tensile deformation and prevent rigid motion were applied at a strain rate of 1/s to a maximum strain of 0.4\%.
\fref{fig:stress_field} shows the transition from elastic stretching to plastic flow in one realization.
Although each grain has a relatively simple response, the collective behavior is complex.
Further details can be found in \cref{frankel2019predicting} and \cref{frankel2022mesh}.

\begin{table}[htb!]
\centering
\begin{tabular}{|c|cr|}
\hline
{\bf Fe} && \\
\hline
Elastic moduli & $C_{11}$ & 204.6 GPa \\
&$C_{12}$ & 137.7 GPa \\
&$C_{44}$ & 126.2 GPa \\
\hline
reference & $\Gamma$ & 1.0 s$^{-1}$ \\
slip rate & & \\
\hline
rate sensitivity & $m$ & 20 \\
exponent &  & \\
\hline
slip resistance & $g_{\alpha}$ & 122.0 MPa \\
\hline
hardening modulus & $H$  & 355.0 MPa \\
\hline
recovery constant & $R$  & 2.9 \\
\hline
\end{tabular}
\begin{tabular}{|c|cr|}
\hline
{\bf Cu} && \\
\hline
Elastic moduli & $C_{11}$ & 180 GPa \\
& $C_{12}$ & 127 GPa \\
& $C_{44}$ &  97 GPa \\
\hline
\hline
{\bf Sb} && \\
\hline
Elastic moduli & $C_{11}$ &  59 GPa \\
& $C_{12}$ &  28 GPa \\
& $C_{44}$ &  26 GPa \\
& $C_{13}$ &  0.64 $C_{12}$ \\
& $C_{33}$ &  1.22 $C_{11}$ \\
& $C_{66}$ &  0.33 $C_{44}$ \\
\hline
\end{tabular}
\caption{Material properties for the polycrystalline elastic and plastic datasets.
For the Fe cubic and Cu/Sb alloy cases the textures of the grains were drawn from a uniform sampling of SO(3), while for the polarized cases the rotation samples were drawn from a distribution  with 0.1 $\times$ 2$\pi$ steradians solid angle dispersion around a pole (001).
The Cu/Sb alloy was in a 9/1 ratio.
Fe (steel) parameters are from \cref{frankel2019predicting}.
Cu and Sb parameters are from \cref{jain2013commentary} and \cref{kammer1972elastic}.
}
\label{tab:material_properties}
\end{table}


\section{Model architecture} \label{sec:arch}

As framed in \eref{eq:A_problem}, the task for the NN model is to approximate the map from the imposed strain $\bar{\strain}(t)$ and the (initial) structure tensor field $\{ \structuraltensor_K \}$ to the average stress $\bar{\stress}(t)$  for the sample.
This homogenization model needs to account for the spatial dependence in the sample volume and,  as stated in \eref{eq:A_equivariance}, it is required to be equivariant so all operations to this mix of input tensors need to be equivariant.
A few methods have been developed to embed equivariant symmetries.
In this work we extend and combine \emph{tensor basis neural networks} (TBNNs) and the equivariant convolutional \emph{tensor field network} (TFN) of Thomas \etal \cite{thomas2018tensor}.

The TBNN \cite{ling2016machine,jones2018machine,jones2022neural} is an equivariant function representation that uses classical representation theorems \cite{spencer1958theory,boehler1987representations,zheng1994theory,rivlin1997stress}; however, as discussed in \sref{sec:representation_problem} it is generally only applicable to single materials, \ie for models of a material point.
Motivated by mixture rules, we use a tensor basis transformation at each cell $K$ to form an approximate stress $\tilde{\sb}_K$ and then equivariant convolution is applied to accommodate local relaxation constrained by continuity/compatibility that is ignored by simple mixture rules.
Following the TFN \cite{thomas2018tensor} we use an equivariant convolution neural network (EqvCNN) that relies on spherical harmonics to form the filters, since these functions are inherently equivariant to rotations and also are expressive.
To accommodate the relative contributions of the cells we also include the cell volume fraction $\nu_K$ in the inputs.
The mesh holding the cell data is encoded as a graph $\graph$ with neighbors and relative distances between cells to facilitate the application of the convolutional filters.
In the overall network, the TBNN provides a learnable reduction/\emph{embedding} of the input data on a per-cell basis, the EqvCNN provides the spatial operations to mix the influence of neighboring cells, and global average pooling emulates the volume average in \eref{eq:A_problem}.

Rather than work with products of tensors in their natural Cartesian basis with arbitrary product rules between tensors of different order, we follow Refs. \cite{thomas2018tensor,cai2023equivariant} and manipulate them in an equivalent form that enables general product rules.
\sref{sec:irrep} gives a brief description of how to handle products of different orders of tensors in a general way through the \emph{irreducible representation} (IR) of group theory \cite{wigner2012group,biedenharn1984angular,tung1985group},
This approach reduces the amount of information produced by general tensor products by omitting the higher order components of the irreducible representations of tensor products.

We begin with the building blocks of equivariant NNs that are suited to the homogenization task, namely local embedding, sample-wise convolution and pooling.
The adaptation of the tensor basis formulation to work on IR vectors is given in \sref{sec:tbnn}.
Pooling is briefly described in \sref{sec:pooling}.
Models for the elastic response of polycrystals are described in \sref{sec:elastic}, while  \sref{sec:inelastic} describes models of anisotropic inelasticity using a novel equivariant recurrent neural network (RNN) and a potential-based neural ordinary differential equation (NODE).
Throughout this section, we will use a superposed hat, such as $\hat{c}$, to indicate learnable NN functions.

\begin{figure}
\centering
\includegraphics[width=0.55\textwidth]{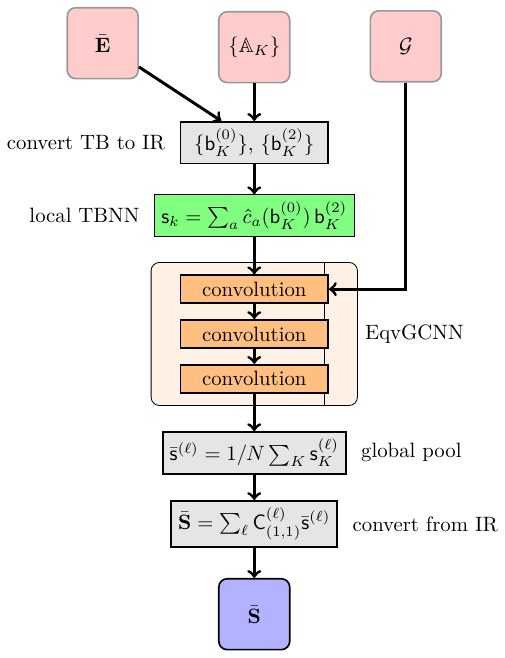}
\caption{Neural network architecture for elastic response.
Inputs are (a) the external strain loading of the sample $\bar{\Eb}(t)$, (b) the structure tensors for each cell $K$, and the graph $\graph$ that provides the cell-to-cell connectivity and cell volume fractions $\nu_K$.
The output is the sample average stress $\bar{\Sb}$.
After converting the scalar invariants and  tensor basis generators of $\bar{\Eb}$ and $\Abb_K$, the local TBNN layer provides an embedding/reduction of these inputs at each cell $K$.
This cell data is then processed by a stack of equivariant convolutions.
The output of the stack is then pooled and this converted to the spatial output $\bar{\Sb}$.
}
\label{fig:eqv_gcnn}
\end{figure}

\subsection{Irreducible representations of tensors} \label{sec:irrep}

An irreducible representation (IR) of a tensor is systematic way of handling tensors of various order and products thereof by decomposing them and flattening them into a collection of vectors.
The map from tensors in their natural representation with basis in $\RE^3$ is accomplished by the Clebsch-Gordon (CG) matrices $\CG{\ell}{\alpha}{\beta}$.
In this notation $\ell \in {|\alpha-\beta|, \hdots, \alpha+\beta}$ is the order of the output IR, and $\alpha$ and $\beta$ are the orders of the input IRs.

For example a symmetric  second order tensor $\Ab$ with basis $\eb_i \otimes \eb_j$ and 6 independent components is decomposed into one 1-vector $\as^{(0)}$ and one 5-vector $\as^{(2)}$
\begin{eqnarray} \label{eq:CGencode}
[ \as^{(0)} ]_i &=& \sum_{j,k=1}^3 [ \CG{0}{1}{1} ]_{ijk} [\Ab]_{jk} \quad i\in\{1\} \\ {} 
[ \as^{(2)} ]_i &=& \sum_{j,k=1}^3 [ \CG{2}{1}{1} ]_{ijk} [\Ab]_{jk} \quad i\in\{1,\ldots,5\} \nonumber
\end{eqnarray}
where $\as^{(0)}$ carries the trace of $\Ab$ and $\as^{(2)}$ holds the deviatoric information.
Likewise tensors of any order can be converted to IR vectors and the transpose of the CG coefficients provide the inverse of the linear change-of-basis map, \eg
\begin{equation} \label{eq:CGdecode}
[\Ab]_{jk} = \sum_{\ell=0,2}  \sum_{i=1}^{2\ell+1}  [ \CG{\ell}{1}{1} ]_{ijk} [ \as^{(\ell)} ]_i   \ .
\end{equation}
The CG transformation applies to both tensors in their natural form and in their IR form where each is a vector in $\RE^{2\ell+1}$ where $\ell$ is order of the order of the IR vector.

Note that our only direct use of the TFN implementation \emph{e3nn} \cite{thomas2018tensor} is to access of the library of CG coefficients.
Further exposition of the fundamentals of IRs is given \aref{app:irrep}.

\subsection{Tensor basis expansion} \label{sec:tbnn}

As we remarked in \sref{sec:representation_problem}, forming a TB expansion across cells or grains is infeasible but we can use a TB layer at the cell level as a form of embedding prior to processing the data on the distance graph $\graph$ with equivariant convolutions.
We chose to apply the TB expansion to the CG encodings of $\Ic_K$ and $\Bc_K$.
This is possible because the CG encoding commutes with the TB expansion.
In the natural Cartesian basis we have
\begin{equation} \label{eq:tb_S}
\tilde{\stress}_K = \sum_a \hat{c}_a(\Ic_K) \Bb_{Ka}
\end{equation}
at each cell $K$ where we ignore invariants and basis generators that cross cells, leaving spatial correlation to the subsequent convolutional layers.
In \eref{eq:tb_S}, $\hat{c}_a(\Ic_K)$ are NN functions of the invariants implemented as a monolithic feed-forward NN with $N_{\Ic}$ inputs and $N_{\Bc}$ outputs.
\begin{eqnarray}
\sum_{j,k} [ \CG{\ell)}{1}{1}]_{ijk}  \tilde{\stress} ]_{jk}
&=& \sum_{j,k} \sum_a [\CG{\ell)}{1}{1}]_{ijk} [ \hat{c}_a(\Ic) \Bb_a ]_{jk} \\
&=& \sum_a \hat{c}_a(\Ic) \sum_{j,k} [\CG{\ell)}{1}{1}]_{ijk} [ \Bb_a ]_{jk} \nonumber\\ {}
[ \mathsf{s}^{(\ell)} ]_i
&=& \sum_a \hat{c}_a(\Ic) [ \bs^{(\ell)}_a ]_i \nonumber
\end{eqnarray}
In effect we obtain one 1-vector $\{ \mathsf{s}_K^{(0)} \}$ corresponding to the approximate pressure in the cell $K$, and one 5 vector  corresponding to the approximate deviatoric stress in the cell $\{ \mathsf{s}_K^{(2)} \}$.

Since the TB expansion is via scalar multiplication and tensor addition, rotation of the inputs commutes with the function and hence the operation is equivariant, \ie
\begin{equation}
\mathsf{s}^{(\ell)}(\Rb)
= \sum_a \hat{c}_a(\Ic) \Ds^{(\ell)}(\Rb) \bs^{(\ell)}_a
= \Ds^{(\ell)}(\Rb) \sum_a \hat{c}_a(\Ic) \bs^{(\ell)}_a  = \Ds^{(\ell)}(\Rb)  \mathsf{s}^{(\ell)}(\Ib)
\end{equation}
in terms of the Wigner $\Ds$ matrices, refer to \aref{app:irrep} and references therein.

\subsection{Equivariant convolution} \label{sec:conv}

Convolution is a form of tensor multiplication between data
$\zs_K = [ \zs^{(\ell)}, \ell=0,..,\ell_\text{max} ]_K $  and a kernel $\fs(\Xb)$ operating over the mesh of cells in the graph $\graph$:
\begin{eqnarray} \label{eq:conv}
\zs^{(\ell)}_K  &=& \Conv_\graph(\zs_K)
= \hat{\fs} \ast \zs
= \sum_{\alpha,\beta} \int \CG{\ell}{\alpha}{\beta} \hat{\fs}^{(\alpha)}(\Xb'-\Xb_K) \zs^{(\beta)}(\Xb') \mathrm{d}^3X'  \\
&\approx&  \sum_{\alpha,\beta} \sum_J \CG{\ell}{\alpha}{\beta} \hat{\fs}^{(\alpha)}(\Xb_J - \Xb_K) \zs^{(\beta)}(\Xb_J) \nu_J  \ ,  \nonumber
\end{eqnarray}
where $\alpha$ is the order of the filter $\hat{\fs}^{(\alpha)}$ and $\beta$ is the order of the input $\zs^{(\beta)}$.
We emphasize that for each convolutional layer this formulation allows for a collection of input vectors with arbitrary orders to be mapped to a collection of output vectors of different orders.
Note unlike \cref{thomas2018tensor}, which was concerned with point clouds, we use the cell volume fraction $\nu_J$ as the integration weight in the summation and use the same filter for all edges of the distance graph $\graph$, i.e. weights are reused throughout the application of the convolution to the graph.

Again we require convolution to be an equivariant operation such that
\begin{equation}
\Ds^{(\ell)} \zs_K^{(\ell)} = \Conv_\graph ( \Ds^{(\beta)} \zs_K^{(\beta)} ) \ ,
\end{equation}
where $\Ds^{(\ell)}$ is the appropriate order Wigner $\Ds$ matrix which effects a change of basis for an IR vector of order $\ell$, refer to \aref{app:irrep}.
As in Thomas \etal \cite{thomas2018tensor}, if the kernel $\hat{\fs}$ takes the special form
\begin{equation}
\hat{\fs}^{(\ell)}(r\rb) = \hat{\varphib}^{(\ell)}(r) \otimes \Ys^{(\ell)}(\rb)  \ ,
\end{equation}
the convolution is equivariant, which relies on the fact that distances are invariant, and directions encoded by the spherical harmonic function to rotate appropriately.
Here $\hat{\varphib}^{(\ell)}(r)$ are learnable NN functions of distance $r = \| \Xb' - \Xb \|$  from the kernel center $\Xb$ and any other point $\Xb'$, and $\Ys(\rb)$ are spherical harmonic functions which only depend on the orientation unit vectors $\rb = (\Xb' - \Xb)/r$.
Note unlike \cref{thomas2018tensor}, which was applied to molecular graphs and used a different $\hat{\varphib}^{(\ell)}$ for each edge, we exploit similarity in the homogenization task and use a unique $\hat{\varphib}^{(\ell)}$ for the entire graph.
This results in a considerable reduction in the number of parameters and no discernible loss of accuracy.

As in \crefs{thomas2018tensor,cai2023equivariant,frankel2022mesh,jones2023deep} we include a self interaction
\begin{equation}
\zs^{(\ell)}_K
= \Conv_\graph(\zs_K)
+  \sum_{\alpha,\beta} \CG{\ell}{\alpha}{\beta} \hat{\ws}^{(\alpha)} \zs_K^{(\beta)} \ ,
\end{equation}
which allows for the extraction of graph Laplacian, grain misorientation and related features.
Note equivariance prevents the addition of a nonzero bias \cite{thomas2018tensor} .
Also the TB embedding can be seen as a special form of a cell-level, self-interaction.

We also follow \cref{thomas2018tensor} and apply nonlinear activation functions to tensors by scaling them by standard activation functions on their norms.
For tensor, $\zs$ in IR form and activation function, $\sigma$, the equivariant activation function is:
\begin{equation}\label{eq:activation}
\sigma_\graph(\zs) = \sigma(||\zs||_2+b)\zs
\end{equation}
where $b$ is a trainable scalar bias.

The complexity of the spherical harmonic expansion is only limited by the hyperparameter $\ell_\text{max}$.
The combinatorics of the general tensor product afforded by the IR products results in a proliferation of terms of different orders and a explosion in the number of parameters needed to define the kernel and self-interaction functions.
For any given application it is unlikely all terms are needed.
Some physical intuition, such as in the TB expansion, can be used to downselect terms, but this is generally not sufficient.
Various techniques exists, such as L$_0$ sparsification \cite{louizos2017learning,fuhg2023extreme}., that can be used to systematically prune the large  combinatorial, product space.
In this work we employed a straightforward greedy algorithm, whereby the least influential products are omitted, the overall network is retrained, and the process is repeated until the loss starts to increase significantly.

\subsection{Pooling} \label{sec:pooling}

The global average pooling we use to go from cell data to sample-level output :
\begin{equation}
\zs^{(\ell)} = \frac{1}{N} \sum_K \zs^{(\ell)}_K ,
\end{equation}
is similar to volume normalized  integration \eref{eq:A_problem}.
Here $N$ is the number of cells being pooled.
This operation is clearly equivariant since it is only requires addition and scalar multiplication.
Note pooling does not change the order $(\ell)$ of the inputs which constrains the output of the preceding convolutional layer if an output of the pooling is required to be of a certain order, as in \fref{fig:eqv_gcnn}.
Local pooling is similar but applied to disjoint patches of the spatial graph to produce a coarser graph.

\subsection{Elastic models} \label{sec:elastic}

The schematic in \fref{fig:eqv_gcnn} depicts a stack of NN layers with equivariant operations that predict the elastic stress of an anisotropic aggregrate.
In this architecture the raw inputs $\bar{\strain}$ and $\{\structuraltensor_K\}$ are first reduced to their joint scalar invariants $\{ \invariants_K \}$, as in  \eref{eq:kam_invar}, and basis generators $\{ \basis_K \}$, as in \eref{eq:kam_basis}.
Then these TB inputs at each node of the distance graph $\graph$ are converted to their equivalent IRs so then can be operated on by the subsequent trainable layers.
The first trainable layer is a TB embedding that operates on each cell independently to produce an approximate cell-wise stress which is fed into the following equivariant convolution stack.
The last layer of the convolutional stack is constrained to produce a field of second order tensors (as represented by one 0th order and one 2nd order IR) which is then pooled and converted to a spatial representation.
This symmetric spatial 2nd order tensor is the estimate of the mean stress for the sample.
In this network exponential linear unit (ELU) activations are used in the TB and equivariant convolutions which are the only trainable aspects.

An alternative architecture which only operates on the scalar invariants to produce a  potential is omitted for brevity since it is subsumed by the inelastic model with the same approach presented in the next section.

\subsection{Inelastic models} \label{sec:inelastic}

The history dependence and internal states of inelasticity requires more complex architectures that account for the causal effects of the imposed strain.

\begin{figure}
\centering
\includegraphics[width=0.45\textwidth]{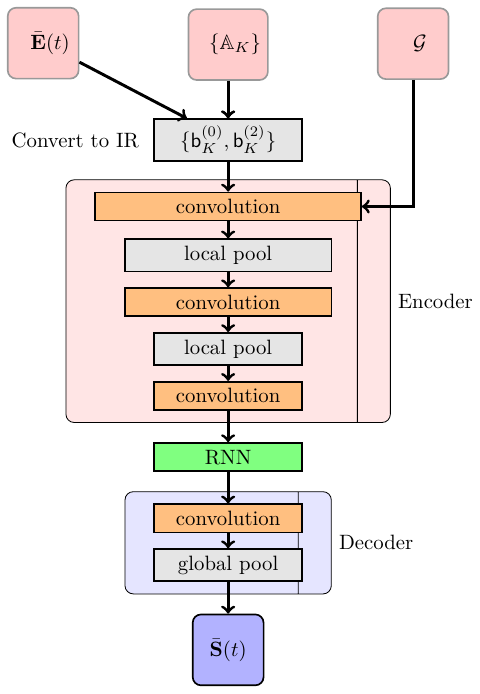}
\caption{An equivariant RNN for inelastic response.
Inputs are (a) the external strain loading of the sample $\bar{\Eb}(t)$, (b) the structure tensors for each cell $K$, and the graph $\graph$ that provides the cell-to-cell connectivity and cell volume fractions $\nu_K$.
Output is the sample average stress $\Sb(t)$ over time $t$.
}
\label{fig:eqv_rnn}
\end{figure}

Recurrent neural networks (RNNs) have been used for this task \cite{frankel2019oligocrystals,bonatti2022importance}, where the hidden states of the RNN model the inelastic state of the material.
The Elman RNN  \cite{elman1990finding} has a simpler form than the more recent gated recurrent unit (GRU) \cite{cho2014learning} and long-term short-term memory unit (LSTM) \cite{hochreiter1997long} RNNs,
but is directly adaptable to equivariant operations.
Specifically, our version of the Elman RNN is
\begin{eqnarray} \label{eq:rnn}
\hs_n &=& \sigma_\graph (\Conv_\graph(\xs_n) + \Conv_\graph(\hs_{n-1})) \\
\ys_n &=& \sigma_\graph (\Conv_\graph(\hs_{n}))  \nonumber
\end{eqnarray}
where $\hs_n$ is the hidden state at step $t_n$ which is initialized to zero at step $t_0$, and $\ys_n$ is the output at at step $t_n$.
Each convolution has an independent kernel and no bias.
The same operations and restrictions can yield equivariant versions of GRU/LSTM.
An encoder-decoder architecture using this RNN is illustrated in \fref{fig:eqv_rnn}.
First the inputs are converted to their IRs.
This field is then processed by a stack of equivariant convolutions and local pooling to reduce it to a coarse field of features.
The RNN propagates this time dependent field which is then processed by another equivariant convolution in the decoder, and finally globally pooled and converted to the spatial output $\bar{\stress}(t)$.
Again, in this network the convolutions were the only trainable aspects and used  ELU activations.

\begin{figure}
\centering
\includegraphics[width=0.45\textwidth]{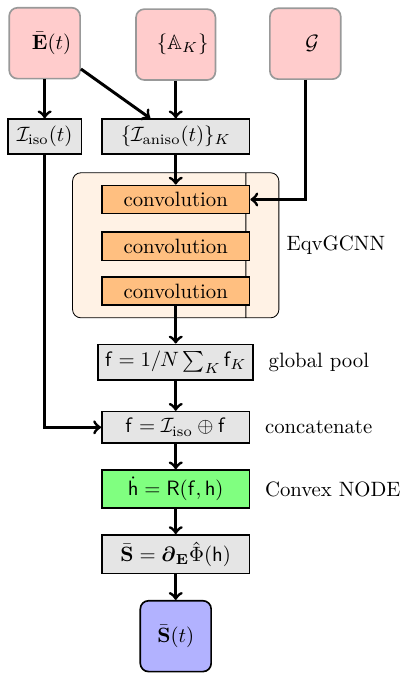}
\caption{ Convex NODE potential GCNN for inelastic response.
Inputs are (a) the external strain loading of the sample $\Eb(t)$, (b) the structure tensors for each cell $K$, and the graph $\graph$ that provides the cell-to-cell connectivity and cell volume fractions $\nu_K$.
The additional latent variables $\hs$ are initialized to zero and the NODE described in detail in \eref{eq:node}.
Output is the sample average stress $\Sb(t)$ over time $t$.
}
\label{fig:scalar_node}
\end{figure}

As an alternative we also implemented a potential-based model  of the evolving inelastic process illustrated in \fref{fig:scalar_node}.
It is along the lines of the model in \cref{jones2022neural} where additional hidden invariants represent the inelastic state in the potential.
Since the scalar invariants and their combinations are trivially equivariant this model is also equivariant.
The hidden features evolve by an (augmented) neural ordinary differential equation (NODE) \cite{dupont2019augmented,chen2018neural} with an input convex \cite{amos2017input} right hand side.
Briefly, the NODE has the update
\begin{equation}
\hs_{n+1} = \hs_n + \Delta\!t \Rs(\fs_n,\hs_n)
\end{equation}
for a simple forward Euler integrator with time-step $\Delta\!t$, where $\hs$ is the augmented hidden state and $\fs$ are the features coming from the convolutional stack in \fref{fig:scalar_node}.
We form the right hand side $\Rs(\xs) = \rs_n$ with an input convex NN (ICNN)
\begin{eqnarray} \label{eq:node}
\rs_1     &=& \sigma( \Ws_0 \xs                + \bs_0  ) \\
\rs_{n} &=& \sigma( \Ws_{n-1} \xs + \Vs_{n-1} \rs_{n-1}  + \bs_{n-1}) \nonumber
\end{eqnarray}
to maintain a degree of regularity in the potential that we subsequently take the gradient of.
Here the weight matrices $\Vs_n$ are required to be positive and the activations $\sigma$ are required to be convex and increasing; the weight matrices $\Ws_n$ and biases $\bs_n$ are unconstrained.
We used \emph{softplus} activations for this NN.
Further details can be found in \cref{amos2017input}.
Since the  isotropic invariants $\tr \bar{\Eb}, \tr \bar{\Eb}^*, \det \bar{\Eb}$  and corresponding basis have no local dependence, unlike the anisotropic invariants and basis which vary across subdomains, we have these isotropic components isotropic components bypass convolution.
The output stress is obtained by taking the gradient of the potential $\Phi$ with respect to strain
\begin{equation}
\stress = \partial_\strain \Phi(\hs) =  \sum_i \underbrace{(\partial_{h_i} \Phi)}_{\hat{c}_i} \, \underbrace{\partial_\strain h_i}_{\Bb_i}
\end{equation}
which effectively forms a basis $\{ \Bb_i =  \partial_\strain h_i \}$ from the set of hidden invariants $\hs$.

\section{Results} \label{sec:results}

As described in \sref{sec:data} we use elastic and elastic-plastic polycrystal SVE datasets demonstrate the proposed architectures.
With the elastic datasets we can explore the representations over a  general strain space, while with the elastic-plastic dataset we can investigate  more complex, history-dependent behavior.
Note all outputs and corresponding errors are for outputs normalized over the training sets.
The inputs, including time, are also scaled to lie in zero to one.

\subsection{Polycrystal elasticity}

For the elastic datasets we use the architecture illustrated in \fref{fig:eqv_gcnn} since there is no history dependence to the stress response, \ie the strain at any point in time and anisotropy field are sufficient to make an accurate prediction.
Specifics of convolutions and other layers  are given in \tref{tab:elasticNN}.

\begin{table}
\centering
\begin{tabular}{|l|c|c|}
\hline
Layer & output & parameters \\
\hline
\hline
Input & $\bar{\Eb}(3,3)$, $\Abb(N_\text{cells},3,3,3,3)$ & 0  \\
Calculate invariants \& basis & $\Ic(N_\text{cells},6)$, $\Bc(N_\text{cells},6,3,3)$ & 0  \\
Convert to irreducible representation & $\mathsf{b}^{(0)}(N_\text{cells},6,1)$, $\mathsf{b}^{(2)}(N_\text{cells},6,5)$  & 0  \\
\hline
Tensor basis embedding & $\mathsf{s}^{(0)}(N_\text{cells},1,1)$, $\mathsf{s}^{(2)}(N_\text{cells},5,1)$ & 74  \\
\hline
Equivariant convolution & $\mathsf{s}^{(0)}(N_\text{cells},1,1)$, $\mathsf{s}^{(2)}(N_\text{cells},5,1)$    & 534  \\
Equivariant convolution & $\mathsf{s}^{(0)}(N_\text{cells},1,1)$, $\mathsf{s}^{(2)}(N_\text{cells},5,1)$ & 528  \\
\hline
Global pooling & $\bar{\mathsf{s}}^{(0)}(1,1)$, $\bar{\mathsf{s}}^{(2)}(5,1)$ & 0 \\
\hline
Convert to Cartesian representation & $\bar{\Sb}(3,3)$  & 0  \\
Output & $\bar{\Sb}(3,3)$ & 0  \\
\hline
\end{tabular}
\caption{Equivariant graph neural network for predicting polycrystalline elastic stress $\bar{\Sb}$ from imposed strain $\bar{\Eb}$ and structure tensor field ${\Abb_K}$.
Dependence on the spatial domain graph $\graph$ and batch have been suppressed for clarity.}
\label{tab:elasticNN}
\end{table}

\subsubsection{Homoanisotropy and heteroanisotropy}
\begin{figure}
\centering
\begin{subfigure}[b]{0.55\textwidth}
\centering
\includegraphics[width=0.9\textwidth]{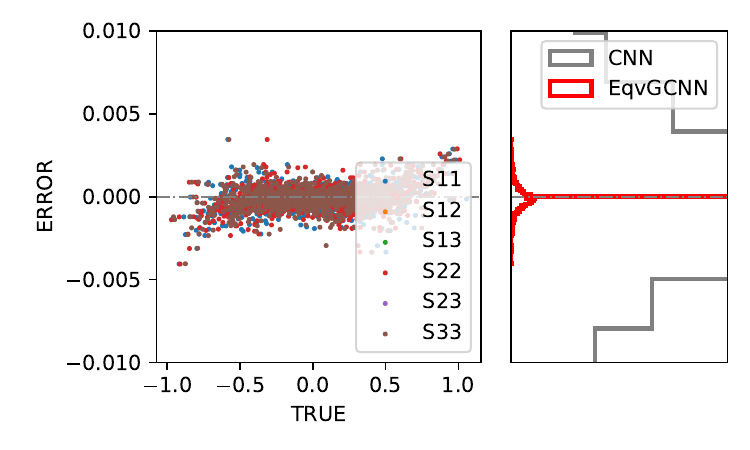}
\caption{single cubic phase, uniform texture}
\end{subfigure}

\begin{subfigure}[b]{0.55\textwidth}
\centering
\includegraphics[width=0.9\textwidth]{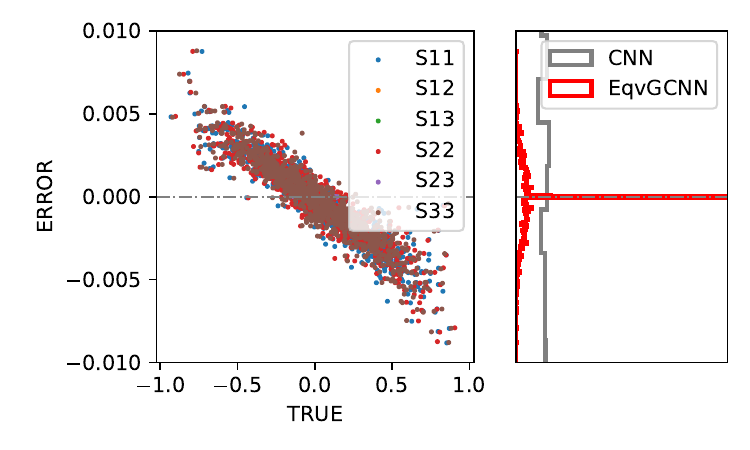}
\caption{binary cubic/tetragonal alloy, uniform texture}
\end{subfigure}

\begin{subfigure}[b]{0.55\textwidth}
\centering
\includegraphics[width=0.9\textwidth]{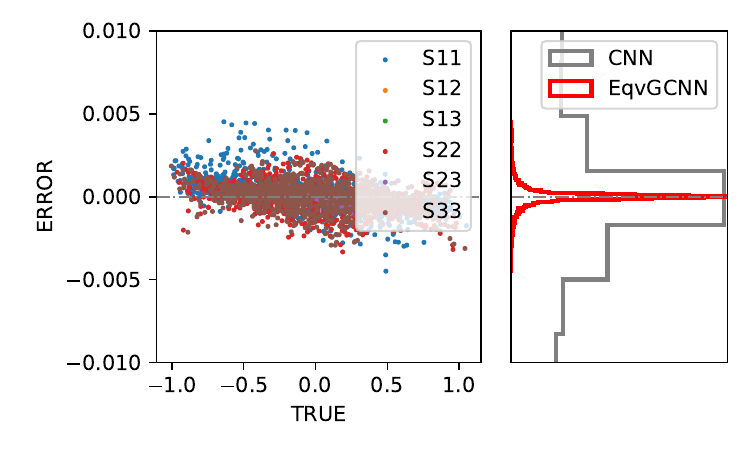}
\caption{single cubic phase, polar texture}
\end{subfigure}

\caption{Crystal elasticity,
Differences with held out data for
(top) cubic, (middle) cubic and (bottem) tetragonal mixture, biased texture orientation samples.
Left: scatter plots of the errors for all stress components, right: distribution of the errors (red) with those of a pixel-based CNN (gray) for comparison.
}
\label{fig:crystal_elasticity}
\end{figure}

\fref{fig:crystal_elasticity} shows the performance of the equivariant network on the three datasets described in  \sref{sec:datasets}.
Although there is some apparent bias to slight under-prediction to the errors for the two more complex cases (alloy and polar texture), overall the predictive quality of the proposed network is excellent with RMSE scores (0.0013, 0.0041, 0.0017) on the held out validation data.
We made a comparison with an analogous pixel-based convolutional NN operating on the field of Euler angles using a GCNN from \cref{frankel2022mesh} with 2 convolutions with 8 filters, global pooling, and 3 dense layers.
The RMSE errors for the pixel-based CNN  (0.077, 0.038, 0.062) are considerably higher, in part due to the complexity in representing the symmetries from the Euler angles.

\subsubsection{Sparsification}
\begin{figure}
\centering
\includegraphics[width=0.45\textwidth]{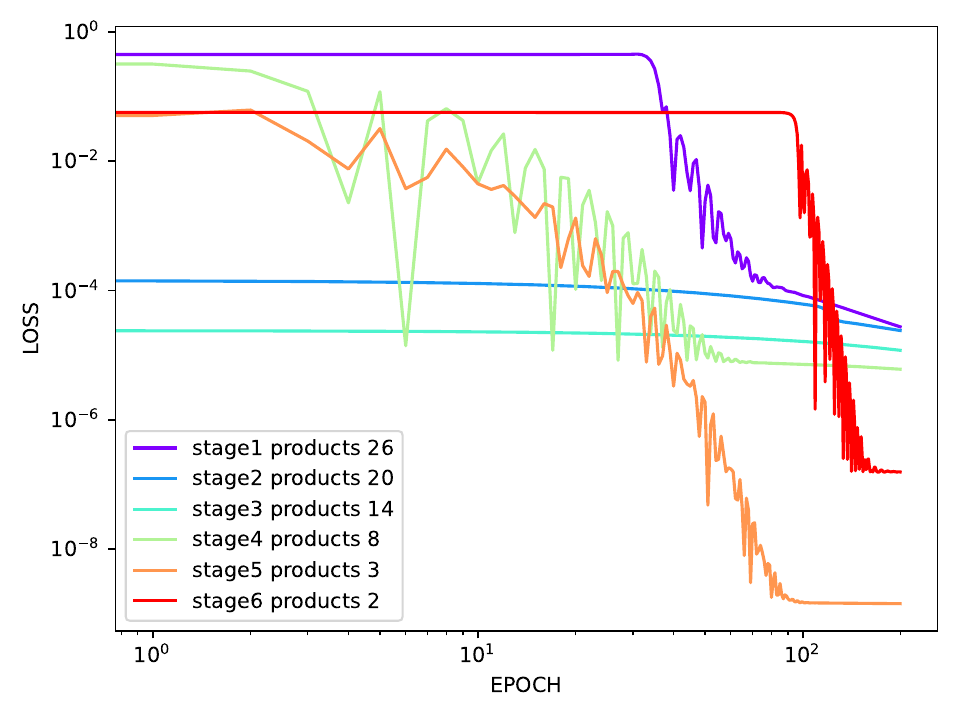}
\caption{Crystal elasticity model sparsification: loss history over greedy pruning stages of the tensor products in the convolutional layers.
}
\label{fig:sparsification}
\end{figure}

\begin{figure}
\centering
\includegraphics[width=0.55\textwidth]{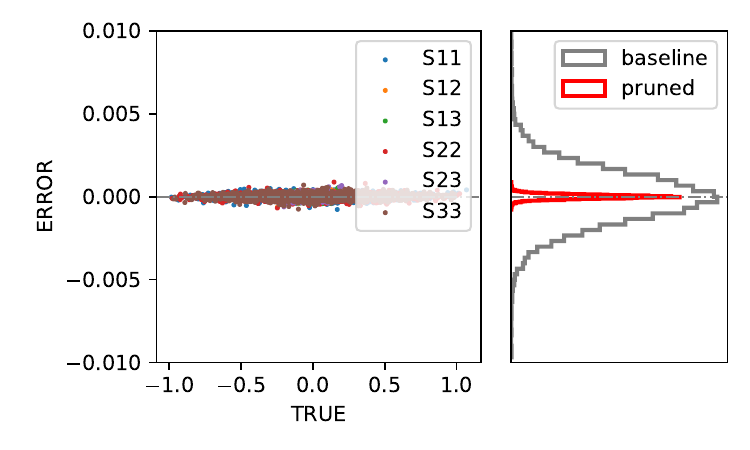}
\caption{Crystal elasticity model sparsification: comparison of pruned and full/baseline models trained to cubic dataset and predicting the polar dataset. Errors shown for the polar dataset (left: all components for the sparse NN, right: distribution of errors for sparse compared to that of full NN).
}
\label{fig:sparsification_validation}
\end{figure}

As mentioned the possible permutations of tensor products for a given input and output tensor field leads to a tremendous number of parameters and corresponding redundancies.
With reference to \eref{eq:conv}, there are a multitude of tensor products are between the trainable filter $\hat{\fs}^{(\alpha)}$ and the selected input tensor $\zs^{(\beta)}$ in IR form that are compatible with a selected input and output.
Hence, we explored the effect of sparsification using the cubic dataset for training.
We employed a straightforward greedy algorithm were we trained the full network  to a plateau in loss, then omitted tensor products with the least significance in terms of the magnitude of the kernel $\hat{\varphi}^{(\ell)}$, and repeated until the accuracy of the network suffered.
\fref{fig:sparsification} shows the sequence of training stages with only the last attempted reduction leading to increase in loss.
The regularity in the loss behavior with the greedy stages lead to easy identification of the (quasi)optimal stopping stage.
\fref{fig:sparsification_validation} shows performance of the network trained to cubic data with the penultimate sparsity on the polar data.
The transferability of the sparse network is remarkable when compared to that of the corresponding un-pruned, full network.
As has been widely observed, the reduction of complexity in the pruned network induces better generalization on held out validation data than the full network which we used as a baseline to explore this transferability.

\subsection{Plastic evolution}

\begin{figure}
\centering
\includegraphics[height=0.35\textwidth]{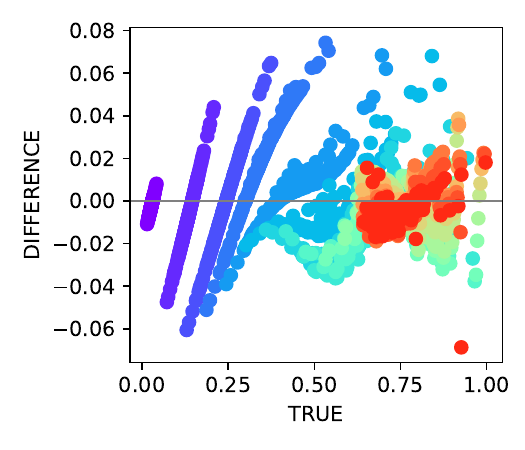}
\includegraphics[height=0.35\textwidth]{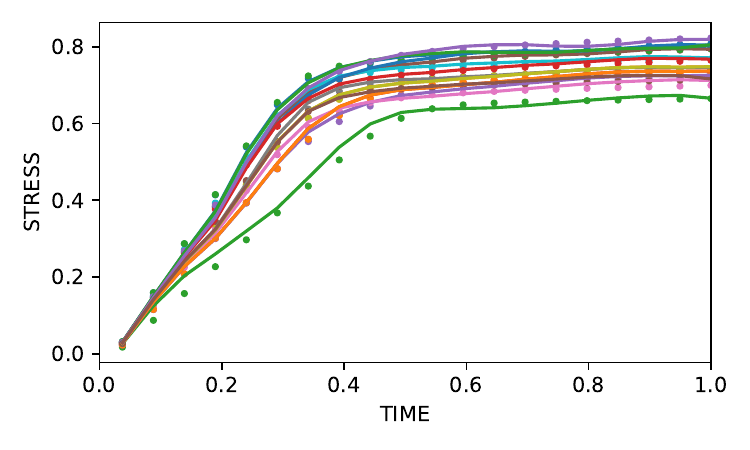}
\caption{Performance of the equivariant RNN on held-out crystal plasticity data: (left) discrepancies with data (colored by time-step red to blue), and (right) predictions of stress evolution plotted as deviation from the mean trend.
Note the left scatter plot shows the errors for the entire held out sample while only randomly selected trajectories are shown on the right.
}
\label{fig:plasticity_rnn}
\end{figure}

\begin{figure}
\centering
\includegraphics[height=0.35\textwidth]{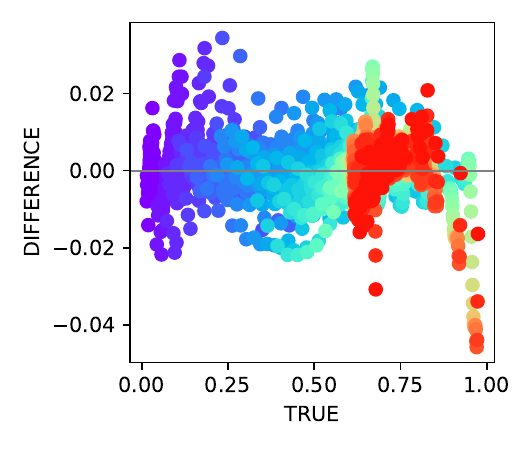}
\includegraphics[height=0.35\textwidth]{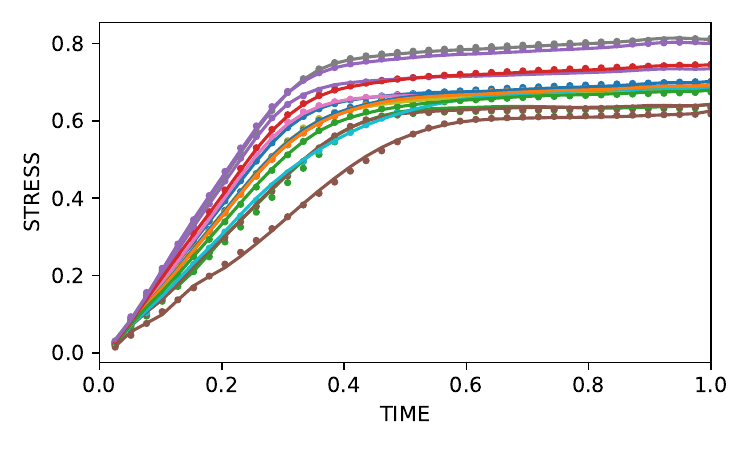}
\caption{Performance of the potential-based NODE on held-out crystal plasticity data: (left) discrepancies with data (colored by time-step red to blue), and (right) predictions of stress evolution plotted as deviation from the mean trend.
Note the left scatter plot shows the errors for the entire held out sample while only randomly selected trajectories are shown on the right.
}
\label{fig:plasticity_pot}
\end{figure}

Using the 3D crystal plasticity simulations of SVEs described in \sref{sec:datasets} we trained an equivariant RNN following the schematic in \fref{fig:plasticity_rnn} and an equivariant potential-based NODE of the type shown in \fref{fig:plasticity_pot}.
The details of the two networks are given in \tref{tab:eqv_RNN} and \tref{tab:scalar_NODE}.
The performance of the RNN-based NN is shown in \fref{fig:eqv_rnn} and the comparable performance of the NODE-based model is shown \fref{fig:scalar_node}, where differences with held out data as a function of the time-step are shown in the left panels and randomly selected trajectories are compared to data in the right panels.
Note in the right panels we plot deviations from mean stress response across dataset ensemble in order to clearly illustrate the level of deviation from the data across the held-out data.
Overall the RMSE for the equivariant RNN was 0.007 and the RMSE for the scalar potential network was 0.001 on the held out data, which compares favorably to the RMSE 0.020 of a non-equivariant GCNN in previous work \cite{frankel2019predicting,frankel2022mesh}.

\begin{table}
\centering
\begin{tabular}{|l|c|c|}
\hline
Layer & output & parameters \\
\hline
\hline
Input & $\bar{\Eb}(N_\text{steps},3,3)$, $\Rb(N_\text{cells},3,3)$ & 0  \\
Convert to IR           & $\mathsf{b}^{(0)}(N_\text{steps},N_\text{cells},1,2)$, $\mathsf{b}^{(2)}(N_\text{steps},N_\text{cells},5,2)$ & 0  \\
\hline
Equivariant convolution & $\mathsf{b}^{(0)}(N_\text{steps},N_\text{cells},1,2)$, $\mathsf{b}^{(2)}(N_\text{steps},N_\text{cells},5,2)$ & 28142   \\
Local pooling           & $\mathsf{b}^{(0)}(N_\text{steps},N'_\text{cells},1,2)$, $\mathsf{b}^{(2)}(N_\text{steps},N'_\text{cells},5,2)$ & 0 \\
Equivariant convolution & $\mathsf{b}^{(0)}(N_\text{steps},N'_\text{cells},1,2)$, $\mathsf{b}^{(2)}(N_\text{steps},N'_\text{cells},5,2)$ & 11270   \\
Local pooling           & $\mathsf{b}^{(0)}(N_\text{steps},N''_\text{cells},1,2)$, $\mathsf{b}^{(2)}(N_\text{steps},N''_\text{cells},5,2)$ & 0 \\
Equivariant convolution & $\mathsf{b}^{(0)}(N_\text{steps},N''_\text{cells},1,2)$, $\mathsf{b}^{(2)}(N_\text{steps},N''_\text{cells},5,2)$ & 11270   \\
\hline
RNN update  & $\mathsf{h}^{(0)}(N''_\text{cells},N_\text{steps},1,2)$, $\mathsf{h}^{(2)}(N''_\text{cells},N_\text{steps},5,2)$ & 2385  \\
\hline
Equivariant convolution & $\mathsf{h}^{(0)}(N''_\text{cells},N_\text{steps},1,2)$, $\mathsf{h}^{(2)}(N''_\text{cells},N_\text{steps},5,2)$ &  5225  \\
Global pooling & $\bar{\mathsf{h}}^{(0)}(N_\text{steps},1,1)$, $\bar{\mathsf{h}}^{(2)}(N_\text{steps},5,1)$ & 0 \\
\hline
Output & $\bar{\Sb}(N_\text{steps},3,3)$ & 0  \\
\hline
\end{tabular}
\caption{Inelastic equivariant RNN.
The encoder reduces the graph from $N_\text{cells} = 25^3$ to $N'_\text{cells} = 10^3$ to $N''_\text{cells} = 4^3$.
The details of the RNN are given in \eref{eq:rnn}.
We have suppressed dependence on the spatial domain graph $\graph$ and batch for clarity.
}
\label{tab:eqv_RNN}
\end{table}

\begin{table}
\centering
\begin{tabular}{|l|c|c|}
\hline
Layer & output & parameters \\
\hline
\hline
Input & $\bar{\Eb}(N_\text{steps},3,3)$, $\Abb(N_\text{cells},3,3,3,3)$ & 0  \\
Calculate invariants & $\Ic_\text{iso}(N_\text{cells},3)$, $\Ic_\Abb(N_\text{cells},3)$ & 0  \\
\hline
Equivariant convolution & $\mathsf{f}^{(0)}(N_\text{cells},N_\text{steps},3,1)$ & 112  \\
Equivariant convolution & $\mathsf{f}^{(0)}(N_\text{cells},N_\text{steps},3,1)$ & 528  \\
\hline
Global pooling & $\bar{\mathsf{f}}^{(0)}(N_\text{steps},3,1)$ & 0 \\
Concatenation & $\Ic_\text{iso}(N_\text{steps},3,1), \bar{\mathsf{f}}^{(0)}(N_\text{steps},3,1)$ & 0 \\
\hline
ODE update  & $\mathsf{h}(N_\text{steps},6)$ & 1581 \\
\hline
Potential & $\Phi(N_\text{steps},1)$ & 6 \\
Gradient & $\partialb_{\bar{\Eb}} \Phi (N_\text{steps},3,3)$  & 0  \\
Output & $\bar{\Sb}(N_\text{steps},3,3)$ & 0  \\
\hline
\end{tabular}
\caption{Inelastic potential-based NODE.
The NODE right hand side NN had 2 layers and 6 augmented states in addition to the 6 time-depedent inputs.
TB embedding not make sense for scalars
We have suppressed dependence on the spatial domain graph $\graph$ and batch for clarity, and dropped the order superscript since all 0-th order IRs.}
Also the conversion of scalars to their IRs is trivial and hence omitted.
\label{tab:scalar_NODE}
\end{table}

\section{Discussion and Conclusion} \label{sec:conclusion}

In this work we have presented a number of effective NN homogenization models using equivariant and tensor basis operations on SVE data, which include sparsity and weight reuse improvements.
These particular architectures illustrate the variety of approaches that can be used to emulate the response of microstructural samples for use in subgrid models, structure-property optimization, and material uncertainty quantification.
In comparison to the graph convolutional neural networks (GCNNs) we have previously employed these models perform better and preserve equivariance and material symmetries exactly.
As others have observed, symmetry preservation allows for configurations that are close up to a symmetry transformation to be treated as similar thus compacting the range of responses that need to be represented.
Although a GCNN is loosely an equivariant network in that the filters are implicitly a function of radial distance only due to treating shells of neighbors alike, we conjecture that our TFN-like networks use of a continuous kernel is a significant contributor to their accuracy.
The use of similar discretization independent constructs, such as neural operators \cite{kovachki2021neural,li2021physics,tripura2022wavelet}, is becoming the state-of-the-art in SciML applications.

There are a multitude of avenues for future work.
Alternative applications abound such as using an equivariant convolutional stack to generate a reduced set of pseudo-structural tensors of various orders directly from the structural tensor field or processing multiple graphs at different scales \cite{jones2023deep} potentially with a U-net structure of equivariant convolutions \cite{ronneberger2015u}.
It may be useful to use transfer convolution layers from the simpler elastic to the inelastic models, with the assumption that the underlying symmetries and features apparent in the elastic deformation phase persist in plastic flow, at least in the early stages.
Predicting the texture induced by deformation is a more challenging task.
Also alternative approaches to sparsification of the tensor products could be employed such as $L^0$ regularization \cite{louizos2017learning,fuhg2023extreme}.
Thomas et al. \cite{thomas2018tensor} introduced the norm based activation functions we use in this work, \eref{eq:activation}. Future work may examine using the TB expansions discussed in this work as an alternative mechanism to design activation functions.
Lastly, new methods for fully sampling and training to the path dependence of the inelastic data have arisen \cite{bonatti2022importance} but this is still a challenge.

\section*{Acknowledgments}

We employed
e3nn (\texttt{https://e3nn.org/}),
TensorFlow (\texttt{https://www.tensorflow.org/}), and
PyTorch (\texttt{https://pytorch.org/}) in the development of this work.

The authors would like to thank Coleman Alleman for providing inciteful feedback on  a draft of this manuscript.
This material is based upon work supported by the U.S. Department of Energy, Office of Science, Advanced Scientific Computing Research program.
Sandia National Laboratories is a multimission laboratory managed and operated by National Technology and Engineering Solutions of Sandia, LLC., a wholly owned subsidiary of Honeywell International, Inc., for the U.S.
Department of Energy's National Nuclear Security Administration under contract DE-NA-0003525.
This paper describes objective technical results and analysis.
Any subjective views or opinions that might be expressed in the paper do not necessarily represent the views of the U.S.  Department of Energy or the United States Government.



\appendix
\section{Irreducible representation of tensors} \label{app:irrep}
\setcounter{equation}{0}
\renewcommand{\theequation}{\thesection.\arabic{equation}}
An irreducible representation (IR) of a tensor is systematic way of handling tensors of various order and products thereof by decomposing them and flattening them into a collection of vectors.

For the purpose of our development, we examine 0th order tensors, such as invariants $\invariants$, and 2nd order tensors, such as the basis generators $\basis$, as examples.
The decomposition of a 0th order tensor (a scalar)  is simply $a^{(0)} = A$, where we use lower case to denote for the coefficients of the IR vector.
A second order tensor can be uniquely decomposed as
\begin{equation}
\Ab =  a \Ib + \perm\, \ab + \devsym\, \ab^\dagger
\end{equation}
where
\begin{align*}
a           &=&& \frac{1}{3} \tr \Ab  \ && a\ \text{is a 1-vector}, \\
\perm\, \ab &=&& \asym(\dev\Ab) && \ab\ \text{is a 3-vector}, \\
\devsym\, \ab^\dagger &=&&  \sym(\dev\Ab) && \ab^\dagger\ \text{contains the remaining 5 components}.
\end{align*}
Here $\dev{\Ab} \equiv \Ab - 1/3 \tr(\Ab) \Ib$, $\asym \Ab \equiv 1/2(\Ab - \Ab^T)$, and $\sym\Ab \equiv 1/2(\Ab' - \Ab'^T)$.
Also $\perm\, \ab = \varepsilon_{ijk} a_k$ with $\varepsilon_{ijk}$ being the third order permutation tensor, and $\devsym \ab  = a_1 (2 \eb_1 \otimes \eb_1 - \eb_2 \otimes \eb_2 - \eb_3 \otimes \eb_3 ) + a_2  (2 \eb_2 \otimes \eb_2 - \eb_1 \otimes \eb_1 - \eb_3 \otimes \eb_3 ) + a_3 (\eb_1 \otimes \eb_2 + \eb_2 \otimes \eb_1) + \ldots$.
The three vectors, $a(\Ab)$, $\ab(\Ab)$, $\ab^\dagger(\Ab)$, form the map, from $\RE^{3 \times 3}$ to $\RE^1 \oplus \RE^3 \oplus \RE^5$.
For a symmetric tensor, such as the strain $\strain$ or stress $\stress$, the 1st order term (the 3-vector) is irrelevant.

\subsection{Clebsch-Gordon matrices}
The Clebsch-Gordorn (CG) transform is a particular version of these maps, generalized to any order tensors.
In fact it provides a decomposition rule for both a single tensor, such as $\Ab = A_{ij} \eb_i \otimes\eb_j$, and the product of tensors, such as  $\ub\otimes\vb$.

To decompose a second order, symmetric Cartesian tensor, $\Ab$ into the irreducible, the CG coefficients, $\CG{\ell}{1}{1}$ for $\ell=0,2$ are needed
\begin{eqnarray} \label{aeq:CGencode}
[ \as^{(0)} ]_i &=& \sum_{j,k=1}^3 [ \CG{0}{1}{1} ]_{ijk} [\Ab]_{jk} \quad i\in\{1\} \\ {} 
[ \as^{(2)} ]_i &=& \sum_{j,k=1}^3 [ \CG{2}{1}{1} ]_{ijk} [\Ab]_{jk} \quad i\in\{1,\ldots,5\} \nonumber
\end{eqnarray}
where $\CG{0}{1}{1}$ is a scaling of the 3$\times$3 identity so $a_0$ is proportional to the trace, and $\CG{2}{1}{1}$ are a set of 5 symmetric traceless tensors that for a basis for all other traceless symmetric tensors.
Note the CG coefficients $\CG{\ell}{\alpha}{\beta}$ are constants, \ie they are  not functions of the encoded tensor.
The CG transform is also orthogonal, so taking the transpose and scaling recovers the Cartesian tensors from the irreducible representations, for example to recover $\Ab$:
\begin{equation} \label{aeq:CGdecode}
[\Ab]_{jk} =  \sum_{i=1}^1  [ \CG{0}{1}{1} ]_{ijk} [ \as^{(0)} ]_i
+  \sum_{i=1}^5  [ \CG{2}{1}{1} ]_{ijk} [ \as^{(2)} ]_i \ .
\end{equation}
where we assume the CG transform is also normalized such that $\CG{\ell}{\alpha}{\gamma} \CG{\ell}{\gamma}{\beta} = \delta_{\alpha\beta}$.

In general, each individual $\CG{\ell}{\alpha}{\beta}$ functions as the bilinear map, $\RE^{2\alpha+1} \times \RE^{2\beta+1} \rightarrow \RE^{2\ell+1}$.
For $\order(\ub)=\alpha$ and $\order(\vb)=\beta$, the irreducible representation components of their tensor product will have order $\ell$ such that $|\alpha-\beta|\leq \ell \leq \alpha+\beta$.
For higher order Cartesian tensors, the decomposition can be applied recursively.
For example, $\CG{\ell}{1}{1}$ provides the map,
\begin{equation} \label{eq:recursiveCG}
\RE^{(3 \times 3) \times 3 \times \hdots}
\rightarrow \RE^{(1+3+5) \times 3 \times \hdots}
= \RE^{1 \times 3 \times \hdots} \oplus \RE^{3 \times 3 \times \hdots} \oplus \RE^{5 \times 3 \times \hdots}
\end{equation}
Each partially decomposed tensor can then be further decomposed by the same process, e.g., use $C^{(j)}_{2,1}$ to decompose the tensor in $\RE^{5 \times 3 \times \hdots}$ into tensors in $\RE^{3 \times \hdots}$, $\RE^{5 \times \hdots}$, $\mathbb{R}^{7 \times \hdots}$.
The Cartesian tensor can be recovered similarly, as in \eref{eq:CGdecode}.
Note that the decomposition is order-dependent but equivalent, so we choose a convention.
For further details refer to \cref{tung1985group}.

\subsection{Wigner $\Ds$ matrices}

The CG matrices provide maps from tensors in their natural form, with basis elements in $\RE^3$, and products of IR vectors to IR vectors.
The Wigner $\Ds$ matrices provide a change of basis operation for IR vectors, analogous to the Kronecker product for tensors in their natural form.
The Wigner $\Ds$ matrices $\Ds^{(\ell)}$ map elements of SO(3) to $(2\ell+1)\times(2\ell+1)$ dimensional (square) matrices, \eg $\Ds^{(0)}(\Rb) = 1$ and $\Ds^{(1)}(\Rb) =  \Rb$.
For example
\begin{equation}
\ms^{(\ell)}(\Rb \Ab) = \Ds^{(\ell)} \ms^{(\ell)}(\Ab)
\end{equation}
is analogous to
\begin{equation}
\Mb(\Rb \boxtimes \Ab) = \Rb \boxtimes \Mb(\Ab)  \ .
\end{equation}
For SO(3), the Wigner-$\Ds$ matrices can be calculated \cite{thomas2018tensor} and the resulting IRs are of dimension, $\ell=1,3,\hdots, 2n+1$.
A representation of SO(3) can in general be block diagonalized such that $V = W_1 \oplus W_2 \oplus  \hdots$, where $W_i$ are proper subspaces of $V$.
In fact a Wigner-$\Ds$ matrix is an \emph{irreducible} representation of proper rotations, $\Ds^\ell: \textrm{SO(3)} \rightarrow GL(V)$, i.e., there is no change of basis, ${\Ds^\ell}'(\Gb) = \Ps^{-1}\Ds^\ell(\Gb) \Ps$, that block diagonalizes  ${\Ds^\ell}'$ for all $\Gb$.
On the other hand, the tensor product of two CG representations is reducible; given two representations, their Kronecker product will decompose according to Clebsch-Gordon (CG) matrices. 

In particular, spherical harmonic tensors $\Ys^{(\ell)}$ are equivariant with respect to the Wigner-$\Ds$ matrices $\Ds^{(\ell)}$ in the sense
\begin{equation}
\Ys^{(\ell)}(\Rb\hat{\rb})
=\sum_{m'} \Ds^{(\ell)}(\Rb) \Ys^{(\ell)} (\hat{\rb})
\end{equation}
where $\hat{\rb}$ is a unit vector, \ie they transform properly with rotation of the coordinate basis.
On the other hand, an affine transformation such as multiplication by a weight matrix $\Ws$ and addition of a bias $\bs$ does not
\begin{equation}
\Ws^{(\ell)} \Ds^{(\ell)} \as^{(\ell)} + \bs^{(\ell)}
\neq
\Ds^{(\ell)} \left[ \Ws^{(\ell)}  \as^{(\ell)} + \bs^{(\ell)}  \right]
\end{equation}
since $\Ws$ does not commute with $\Ds$ and the bias $\bs$ is not affected by the rotation of $\as$.

For more details refer to \crefs{biedenharn1984angular,tung1985group,cornwell1997group}.

\end{document}